\documentclass[useAMS,usenatbib]{mnras}
\usepackage{threeparttable}
\usepackage{amsmath}
\usepackage{graphicx}
\usepackage{grffile}
\usepackage{hyperref}
\usepackage{txfonts}
\usepackage{natbib}
\usepackage{bm}
\usepackage{color}

\title[A Changing Look Event in the NGC~3516]{Multi-Wavelength Monitoring and Reverberation Mapping of a Changing Look Event in the Seyfert Galaxy NGC~3516}

\author[V. L. Oknyansky et al.]{V. L. Oknyansky,$^{1}$\thanks{E-mail: oknyan@mail.ru}
M. S. Brotherton, $^{2}$
S. S. Tsygankov,$^{3,4}$
A. V. Dodin,$^{1}$
D. -W. Bao, $^{5}$
\newauthor
B. -X. Zhao, $^{6}$
P. Du,$^{5}$
M. A. Burlak,$^{1}$
N. P. Ikonnikova,$^{1}$
A. M. Tatarnikov,$^{1}$
\newauthor
A. A. Belinski,$^{1}$
A. A. Fedoteva,$^{1}$
N. I. Shatsky,$^{1}$
E. O. Mishin,$^{1}$
S. G. Zheltouhov,$^{1}$
\newauthor
S. A. Potanin,$^{1,7}$
J. -M. Wang, $^{5,8}$
J. N. McLane, $^{2}$
H. A. Kobulnicky, $^{2}$
D. A. Dale, $^{2}$
\newauthor
T. E. Zastrocky, $^{2,9}$
J. Maithil, $^{2}$
K. A. Olson, $^{2}$
C. Adelman, $^{2,10}$
Z. Carter, $^{2,11}$
\newauthor
A. M. Murphree, $^{2,12}$
M. Oeur, $^{2,13}$
S. Schonsberg, $^{2,14}$
T. Roth $^{2,15}$
\\
$^{1}$Sternberg Astronomical Institute, M.V. Lomonosov Moscow State University,  119234, Moscow, Universitetsky pr-t, 13, Russia\\
$^{2}$Department of Physics and Astronomy, University of Wyoming, Laramie, WY 82071, USA\\
$^{3}$Department of Physics and Astronomy, FI-20014 University of Turku, Turku, Finland\\
$^{4}$Space Research Institute of the Russian Academy of Sciences, Profsoyuznaya Str. 84/32, Moscow 117997, Russia\\
$^{5}$Key Laboratory for Particle Astrophysics, Institute of High Energy Physics, Chinese Academy of Sciences, 19B Yuquan Road, Beijing 100049, China\\
$^{6}$Shanghai Astronomical Observatory, Chinese Academy of Sciences, Shanghai 200030 China\\
$^{7}$Faculty of Physics, Moscow M.V. Lomonosov State University, Leninskie gory 1, Moscow, 119991, Russia\\
$^{8}$School of Astronomy and Space Science, University of Chinese Academy of Sciences, 19A Yuquan Road, Beijing 100049, China\\
$^{9}$Physics and Astronomy Department, Regis University, Denver, CO 80212, USA\\
$^{10}$Department of Physics \& Astronomy, Poly Pomona, Pomona, CA 91768, USA\\
$^{11}$Department of Physics and Astronomy, Trinity University, San Antonio, TX 78212, USA\\
$^{12}$Department of Physics, Rhodes College, Memphis, TN 38112, USA\\
$^{13}$Department of Physics and Astronomy,  State Long Beach, Long Beach, CA 90840, USA\\
$^{14}$Department of Physics and Astronomy, University of Montana, Missoula, MT 59812, USA\\
$^{15}$Department of Physics \& Astronomy, California State University, Sacramento, CA 95747, USA
}
\date{Accepted XXX. Received YYY; in original form ZZZ}

\pubyear{2021}

\begin{document}
\date{Received ... Accepted ...}
\pagerange{\pageref{firstpage}--\pageref{lastpage}}
\maketitle
\label{firstpage}

\begin{abstract}%
We present the results of photometric and spectroscopic monitoring campaigns of the changing look AGN NGC~3516 carried out in 2018 to 2020 covering the wavelength range from the X-ray to the optical. The facilities included the telescopes of the CMO SAI MSU, the 2.3-m WIRO telescope, and the XRT and UVOT of $Swift$. We found that
NGC~3516  brightened to a high state and could be classified as Sy1.5 during the late spring of 2020. We have measured time delays in the responses of the Balmer and He II $\lambda4686$~lines to continuum variations. In the case of the best-characterized broad H$\beta$ line, the delay to continuum variability is about 17 days in the blue wing and is clearly shorter, 9 days, in the red, which is suggestive of inflow.  As the broad lines strengthened, the blue side came to dominate the Balmer lines, resulting in very asymmetric profiles with blueshifted peaks during this high state.
During the outburst the X-ray flux reached its maximum on 1 April 2020 and it was the highest value ever observed for NGC~3516 by the {\it Swift} observatory. The X-ray hard photon index became softer, $\sim$1.8 in the maximum on 21 Apr 2020 compared to the mean $\sim$0.7 during earlier epochs before 2020. We have found that the UV and optical variations correlated well (with a small time delay of 1--2 days) with the X-ray until the beginning of April 2020, but later, until the end of Jun. 2020, these variations were not correlated. We suggest that this fact may be a consequence of partial obscuration by Compton-thick clouds crossing the line of sight.

\end{abstract}






\section{Introduction}

The spectral variability of NGC~3516 was first discovered  by \cite{as68} when they compared their 1967 spectrum and spectrophotometric results (for 1965-1966) from \cite{Dibay1968} with  earlier data from  \cite{se43}. That was not only the first detection of spectral variability for Seyfert galaxies but also the first example of an Active Galactic Nucleus (AGN) changing its spectral type - a Changing-Look (CL) AGN. That happened a bit earlier than the continuum variations in the Seyfert galaxy NGC~4151 were reported by \cite{Fitch1967} and, therefore, \cite{as68} suspected the forbidden lines to have changed dramatically, and the continuum and Balmer lines not to be short-term variable, although the reverse is accepted as true today. The discovery of the H${\alpha}$ variability for NGC~3516 and, at the same time, the first reverberation mapping (RM) for line variations relative to the continuum (for AGN in general) was  done in 1973 by \cite{Ch1973}. To date, the CL events were found in tens of AGNs including QSOs \citep[see e.g.,][]{Macleod2016, Runco2016}. A detailed study of a CL AGN can be very useful for the understanding of the physics and evolution of AGNs.

NGC~3516 was investigated very intensively during the past half century. The studies included high-cadence optical photometry \citep{ld93}, spectral monitoring  \citep{wa93, fa16,sh19}, X-ray observations \citep{nd16,bu17} and IR photometry \citep{Koshida2014}. The first RM (with the use of some statistical analysis) for NGC~3516 was done by \cite{wa93} who found time delays of about 15 and 7 days between the H$\alpha$ and H$\beta$ variations and the continuum changes, respectively.

The velocity-resolved RM program conducted in 2007 \citep{Denney2009} for the H$\beta$ emission line demonstrated that the highest positive velocities exhibited the shortest lags, with the lags steadily growing toward negative velocities. The situation appears to have changed to the opposite in 2012 \citep{dr18}. The delay for different velocity bins in the emissionline profile varied from a few days to 18 days. This behaviour could be interpreted as a signature of infalling gas motion in 2007 and outflowing in 2012, respectively. The RM relative to X-ray and UV variations yielded controversial results. The first conclusion that the delay between the X-ray flux variations and those of the optical is about 100-200 days \citep{Ma02} was not subsequently confirmed by later studies, whereas a delay of 2 days between X-ray and UV variability seemed more likely \citep{nd16,bu17}.

During the last 8 years NGC 3516 was in a very low state with some re-brightening at the end of 2015 and the first half of 2016. In 2014 a CL event was detected when the broad H$\beta$ was quite faint \citep{sh19} and an additional CL event was observed in 2016 when broad emission lines were in the high state again \citep{Oknyansky2020b}. From the second half of 2016 till the end of 2018 the object was in the low state with broad emission lines being very weak. In our previous study (\citet{Ilic2020} = Paper I) we reported on the awakening of the object after several years of being in a very low state. We found a brightening of the continuum with a maximum at the end of Nov. 2019. At the beginning of Dec. 2019  the broad double-peaked components of H$\beta$ and H$\alpha$ were detected and they were more prominent than in 2018 \citep{sh19}. Also strong high-ionisation coronal lines like [\ion{Fe}{x}]$\lambda$6374 were present in the spectra and suspected to be variable. Preliminary results of our research on the flare and new CL event in NGC3516 were mentioned by \cite{Oknyansky2020a}. Here we present a continuation of our research based on newer and more extensive data for NGC~3516.  Such consistently variable objects as NGC~3516 straddling the line between different classes represent special laboratories to study CL events.

\section{Observations, instruments and data reduction}

We started photometric (\textit{UBVRcIc}) monitoring of NGC~3516 in September 2019 in order to check if the object was in the low state as reported by \cite{sh19}. As soon as we detected a brightening at the end of November 2019 we started spectral observations (from 7 December 2019) with the 2.5-m telescope of the Caucasian Mountain Observatory (CMO) of the Sternberg Astronomical Institute (SAI) of Moscow State University (MSU). Preliminary results of the research were published in Paper I. Here we present the next part of the study based on intensive photometric and spectral observations  from December 2019 to July 2020. We also report new {\it Swift} X-ray, UV/optical data obtained  in Feb.--Jun. 2020. The data are complemented by high-cadence spectroscopy obtained with the 2.3-m Wyoming Infrared Observatory (WIRO) telescope in 2018--2020. The details of the observations and data reduction as well as relevant references are given in the next subsections, whereas the results are presented and discussed in the following section.

\subsection{Optical photometry}

\begin{figure}
   \includegraphics[width=\columnwidth]{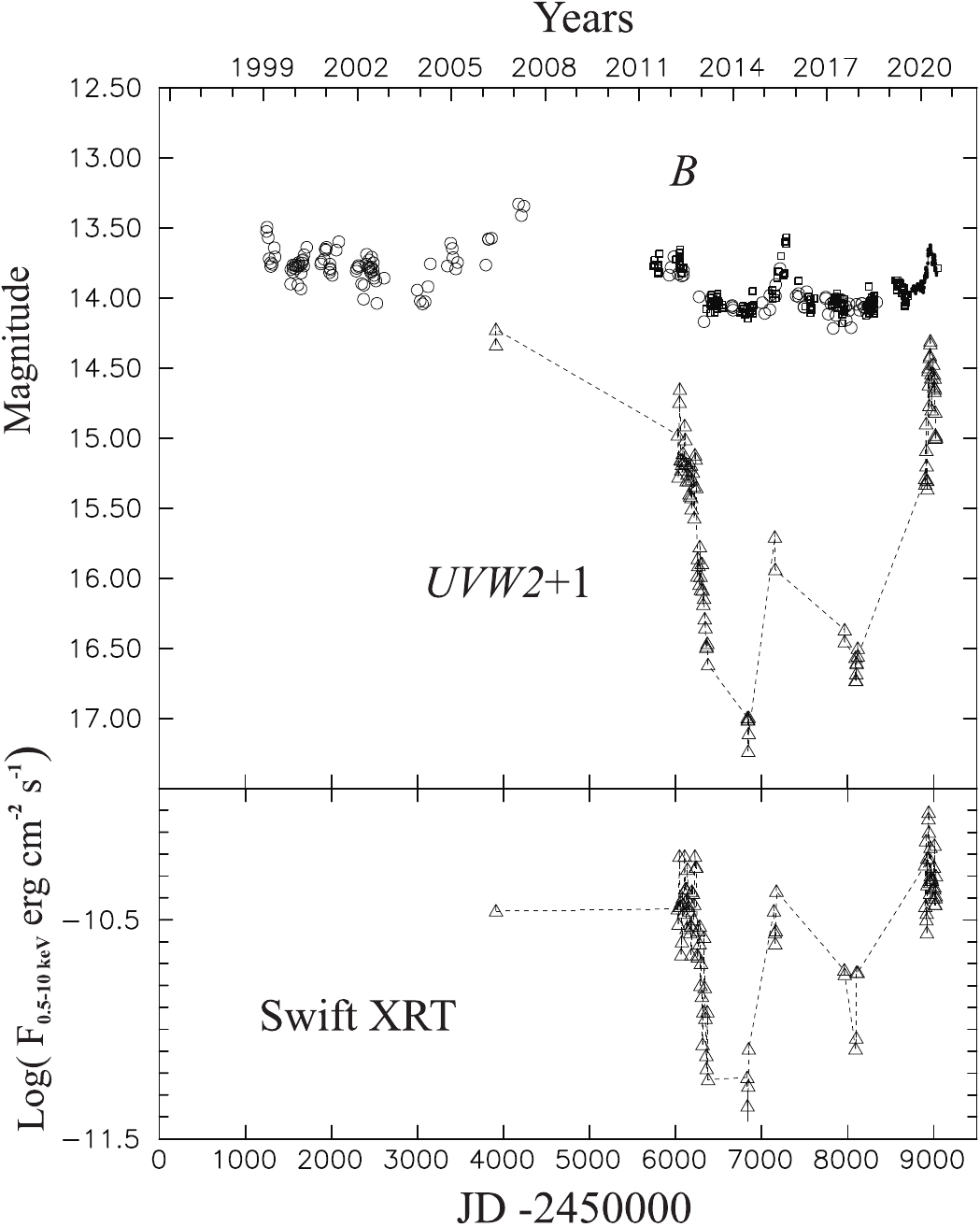}
   \caption{Multi-wavelength observations of NGC~3516 in 1999-2020.  {\it Top Panel}: Light curves in the $B$ and $UVW2$ bands for an aperture of 10 arcsec. The large open circles denote data from  \protect\cite{sh19}, the open boxes stand for the data from the SAI Crimea campaign  and are shown for continuity  (see \protect\cite{ld93} and Paper I for details). The points represent the estimates (nightly means) obtained with RC600 (partly presented in Paper I). The open triangles are the $UVW2$ data obtained with  {\it Swift}/UVOT.
   {\it Bottom Panel}: The {\it Swift}/XRT 0.5--10 keV  X-ray flux (in ergs cm$^{-2}$ s$^{-1}$).
    }
    \label{fig1}
\end{figure}

\begin{figure}
	\includegraphics[width=\columnwidth]{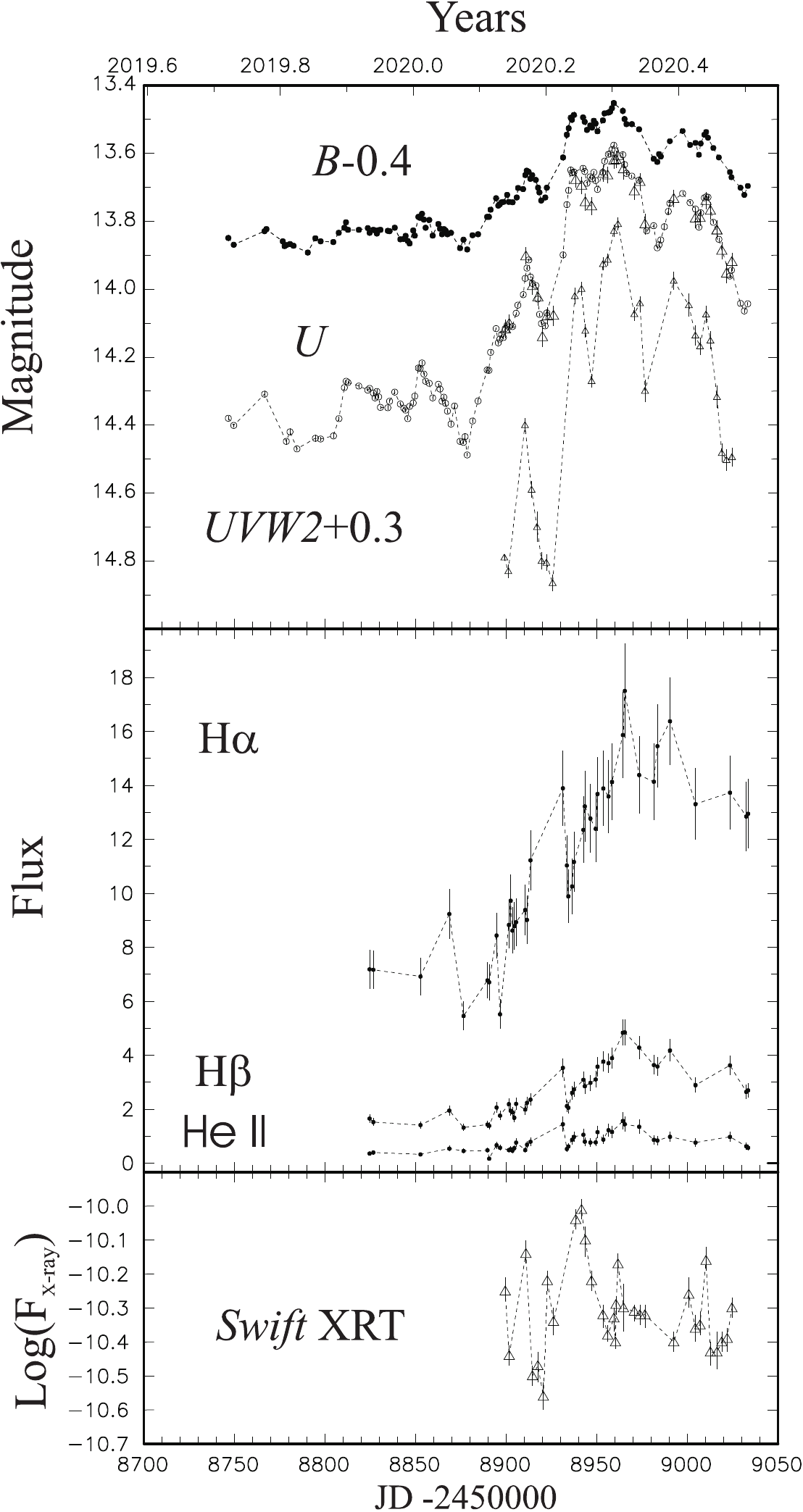}
    \caption{Multi-wavelength observations of NGC~3516 shown for 2019-2020 only.  {\it Top Panel:} Light curves in $BU$ (an aperture of 6.7 arcsec) and $UVW2$ (an aperture of 10 arcsec). The nightly mean $B$ (filled circles) and $U$ (open circles) estimates obtained with RC600. The $U$ data obtained with {\it Swift}/UVOT and reduced to the same system and aperture as the RC600 data  (triangles).
     {\it Middle Panel:}
 Variations of the fluxes (in units 10$^{-13}$ erg s$^{-1} $cm$^{-2}$) of H$\alpha$, H$\beta$ and He II from the spectral data obtained with the 2.5-m telescope of CMO SAI MSU (see text).
 {\it Bottom Panel:}   The {\it Swift}/XRT 0.5--10 keV  X-ray flux (in erg cm$^{-2}$ s$^{-1}$) -- (triangles).
   }
    \label{fig2}
\end{figure}

We obtained optical \textit{UBVRcIc} CCD data with RC600, a new 60-cm automated telescope located at the CMO, for nearly 10 months since September 2019 (a total of 126 nights, with a cadence of about 2 days). The telescope, photometer, and the photometry methods were described by \cite{Berdnikov2020} and also in Paper~I. Comparison stars from \cite{ld93} were used. During this more than 9 month period, the data were obtained mostly under photometric conditions, several exposures (2-6) were made for each band, with typical exposure times varying from 300-360~s for the $U$ band to 20-30~s for $B$. We used an aperture of 6.7 arcsec (diameter). The consistency of the individual multiple magnitudes measured each night, usually three, was usually not worse than 0.01 mag, however the error could be few times bigger in case of bad conditions or during bright Moon light contamination. The $UB$ light curves are presented in Fig.~\ref{fig1}-\ref{fig2} and Table ~\ref{Tab1d}. In Fig.~\ref{fig1} we also show the already published optical $B$ photometry since 1999 (see \cite{sh19} and Paper I for details).

\begin{table}
\centering
\caption{Results of optical photometry with the RC600 CMO in $UB$ bands. ( Note:  The full version of this table is available in its entirety in machine-readable form.)}

\begin{tabular}{ccc} \hline
   J.D.-2450000& mag  & Band\\
   \hline
8746.54 & 14.249 $\pm$ 0.010 & $B$\\
8746.54 & 14.380 $\pm$ 0.006 & $U$\\
8749.53 & 14.401 $\pm$ 0.006 & $U$\\
8749.55 & 14.269 $\pm$ 0.004 & $B$\\
8766.55 & 14.309 $\pm$ 0.010 & $U$\\
...&  ...&...\\ \hline


 \\

\end{tabular}

 \label{Tab1d}
\end{table}

\subsection{{\it Swift} Optical, Ultraviolet, and X-ray observations}

The {\it Neil Gehrels Swift Observatory} \citep{2004ApJ...611.1005G} observed NGC~3516 in 2006 but only since 2012 it started semi-regular monitoring. Some of these data (obtained in 2006--2017) were published in \cite{bu17} and Paper~I. In addition to the data available in the archive, we applied for a new TOO program starting from 19 Feb 2020 and continuing until the end of June 2020 (a total of 32 new dates), with a cadence of 3 days (however some of the scheduled observations were not executed and so the real mean sampling was just about 4 days).  All the data were reduced uniformly with the most recent versions of the software and calibration files. 

The {\it Swift} Ultraviolet/Optical Telescope (UVOT) observes the source in different filters ($V$, $B$, $U$, $UVW1$, $UVW2$, $UVM2$) simultaneously with the XRT telescope. The image analysis has been done following the procedure described on the web-page of the UK Swift Science Data Centre. Photometry was performed using the procedure {\sc uvotsource} from the {heasoft} package. The source and background apertures were chosen with radii of 5 and 10 arcsec, respectively.

The XRT telescope \citep{2005SSRv..120..165B} observed NGC~3516 in photon counting (PC) and windowed timing (WT) modes, depending on
target brightness. All the new observations were made only in PC mode. The spectrum obtained in each observation was prepared using the online tools provided by the UK Swift Science Data Centre \citep[\url{http://www.swift.ac.uk/user_objects/};][]{2009MNRAS.397.1177E}. The method used was the same as described in \cite{Oknyansky2017a, Oknyansky2019a} and Paper I.
The light curves in X-ray {\it Swift}/XRT) and ultraviolet  ({\it Swift}/UVOT/$UVW2$)  are presented in Fig.~\ref{fig1} (2006--2020) and Fig.\ref{fig2} (2020). The $U$  data obtained with {\it Swift}/UVOT presented in Fig.\ref{fig2} are reduced to the same system and aperture as the RC600 data using a linear least squares regression. Obtained data give us opportunities to trace the source evolution on long and short timescales.

\subsection{Spectral observations  at CMO SAI MSU}

\begin{figure}
	\includegraphics[width=\columnwidth]{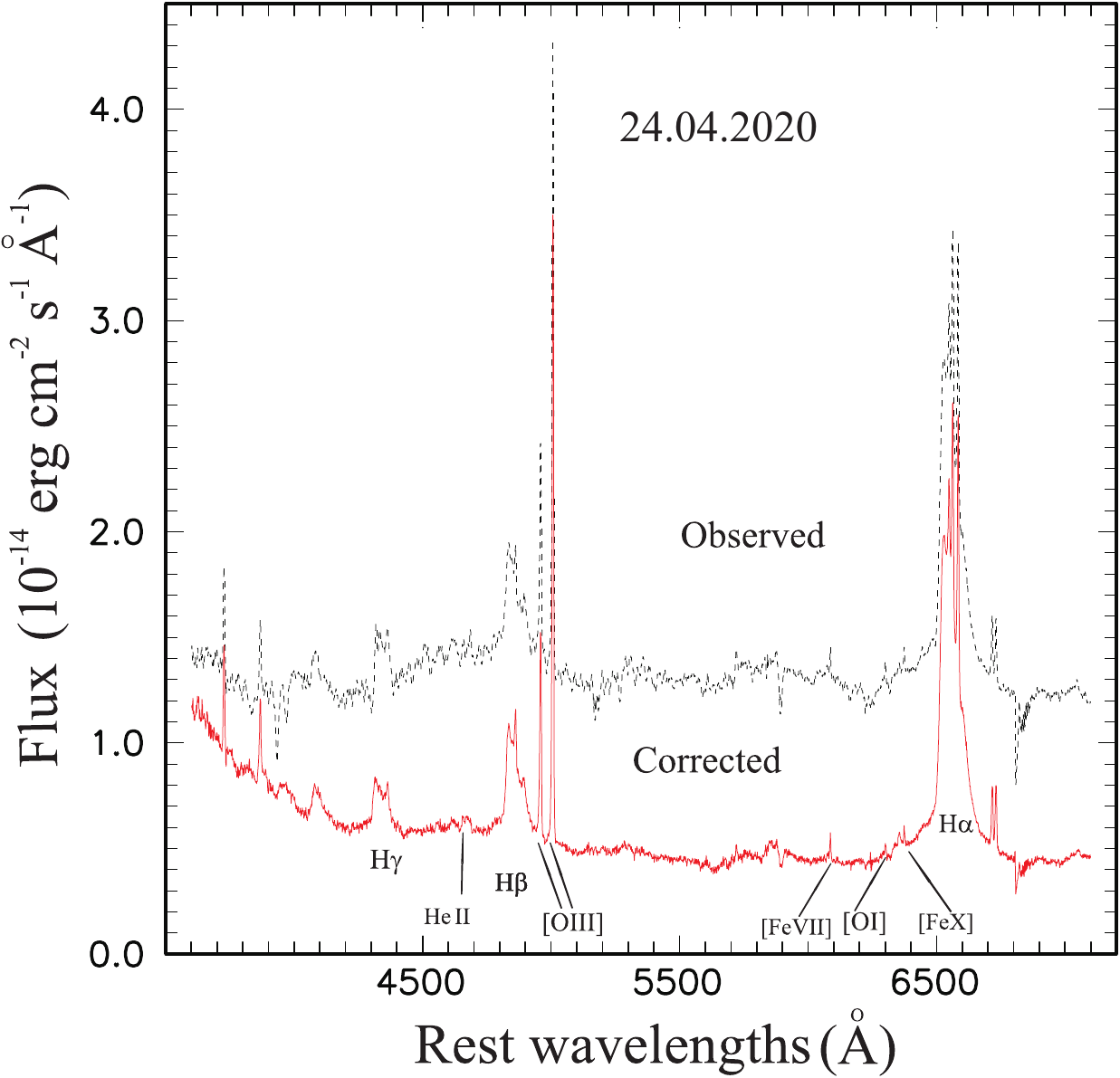}
    \caption{The isolated nuclear non-stellar spectrum (solid line) of NGC~3516 for 24.04.2020 obtained by subtracting the host galaxy spectrum from the observed (dashed line). See text for details.}
    \label{fig3}
\end{figure}

\begin{figure*}
	\includegraphics[scale=0.9,angle=0,trim=0 0 0 0]{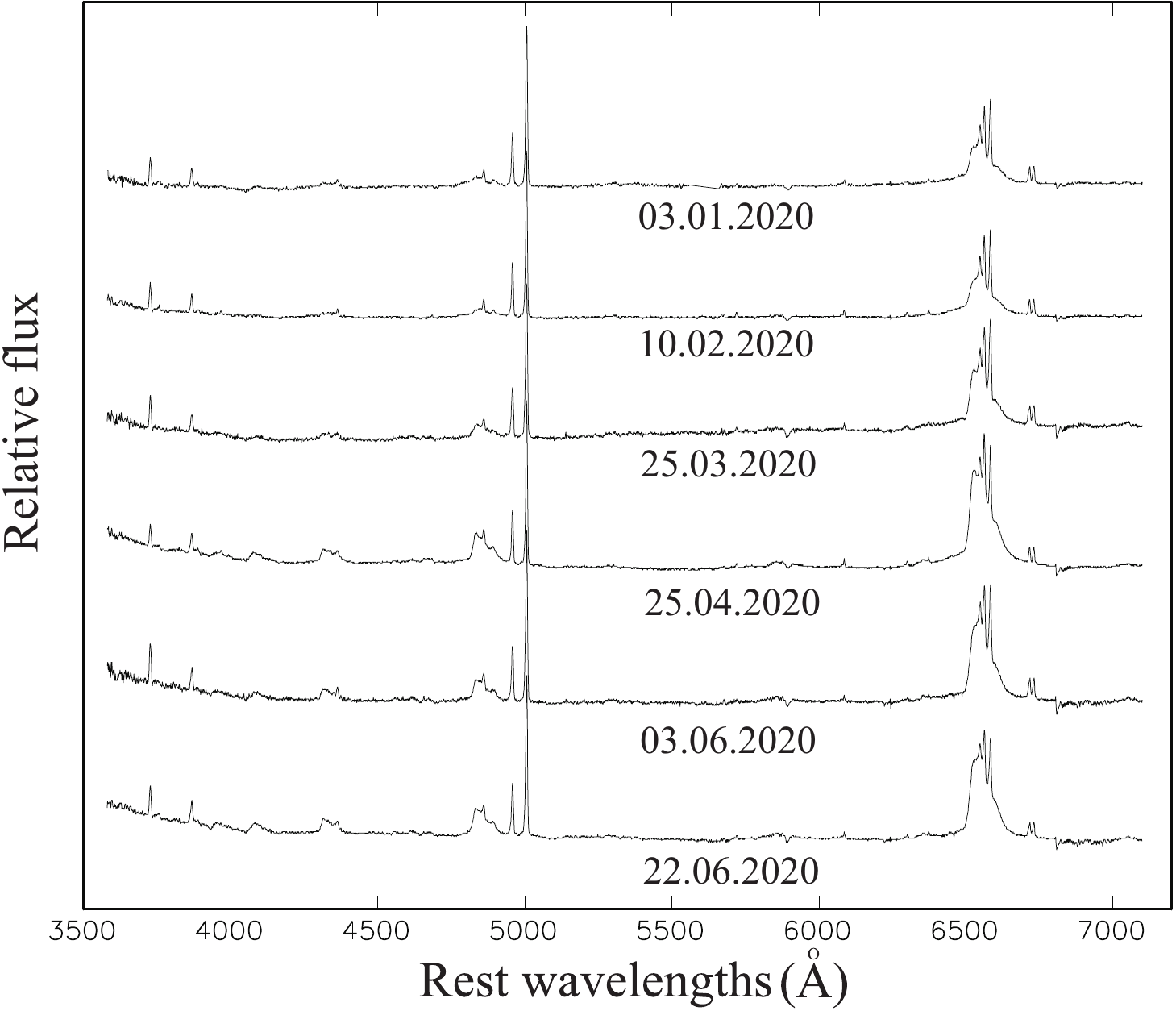}
    \caption{The selected host-galaxy subtracted spectra, obtained in the CMO for those spectra covering the total wavelength range. All spectra are normalized to the [O III]$\lambda$5007 intensity and shifted for comparision.}
    \label{somecmo}
\end{figure*}

\begin{figure}
	\includegraphics[width=\columnwidth]{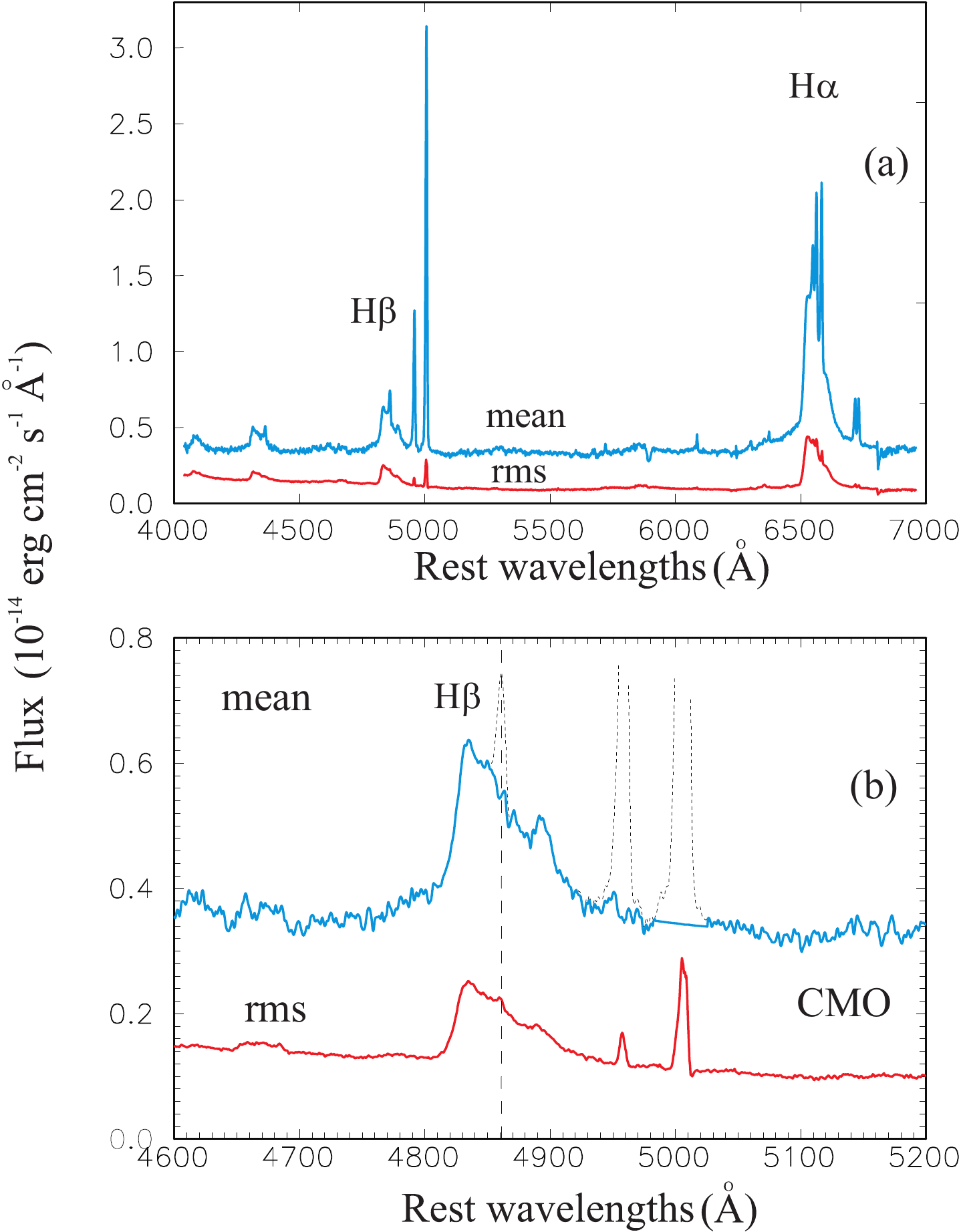}
    \caption{Mean and rms spectra (CMO) in the rest-frame for the 
full spectral region (a) and just for the H$\beta$ region (b). In the bottom 
panel the dotted lines are the narrow H$\beta$ and [OIII]$\lambda$4959,5007 in the 
mean spectrum (see more details in the text), while the dashed line 
indicates 4861 \AA~in order to illustrate the asymmetric H$\beta$ profile more 
clearly.
        }
   \label{rmscmo}
\end{figure}

\begin{figure}
	\includegraphics[width=\columnwidth]{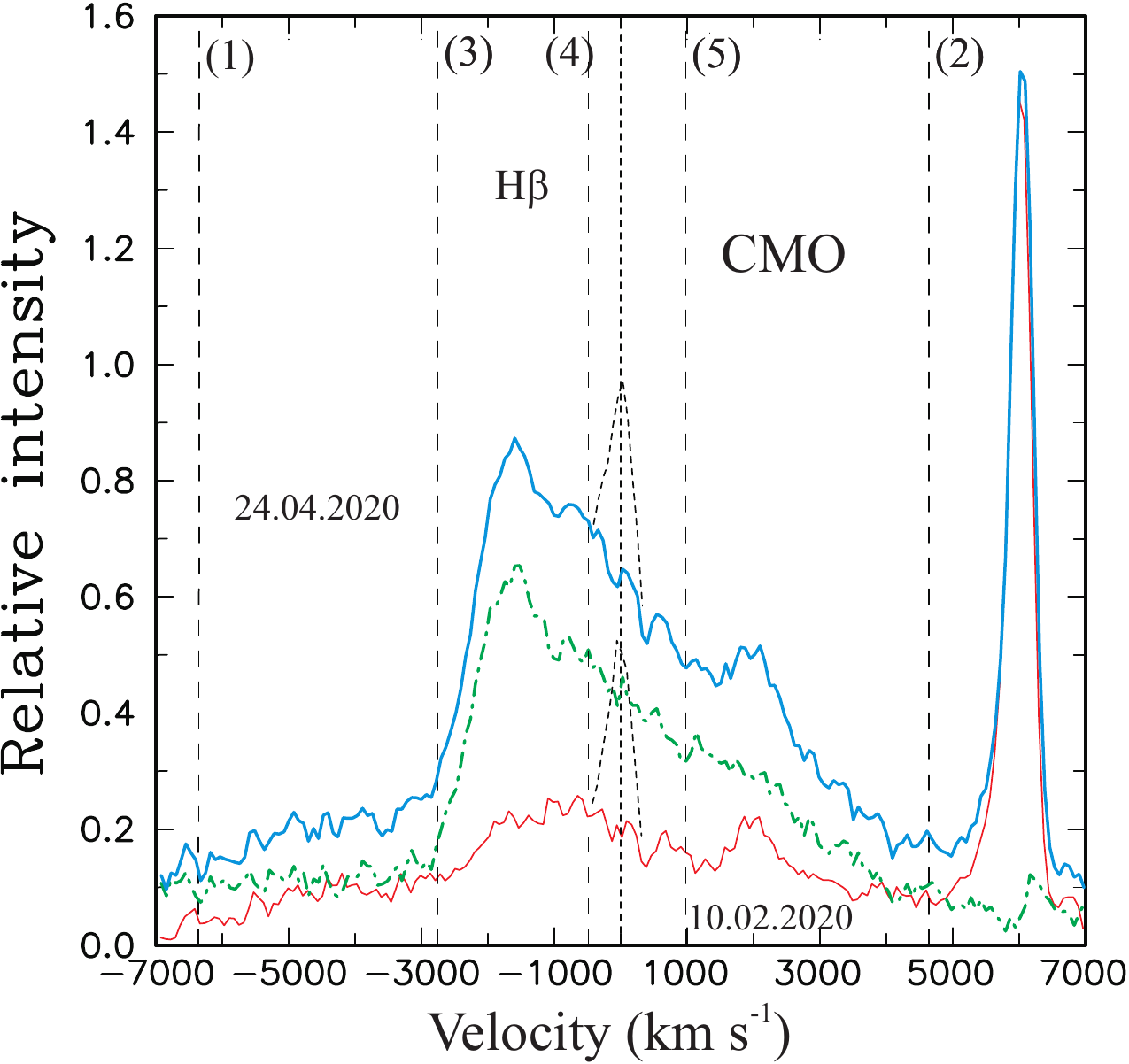}
    \caption{Profiles of broad H$\beta$ for the maximal (24.04.2020) and  minimal (10.02.2020) states (with the narrow components removed) showing intrinsic variations of the line. The difference between the maximal and  minimal states is shown by the dash-dotted line. The vertical dashed lines (1-5) mark the regions for integration of I(H$\beta$)~(1-2), I(H$\beta$(Blue))~(3-4) and I(H$\beta$(Red))~(5-2). See text for details.}
    \label{fig4}
\end{figure}

\begin{figure*}
	\includegraphics[scale=0.9,angle=0,trim=0 0 0 0]{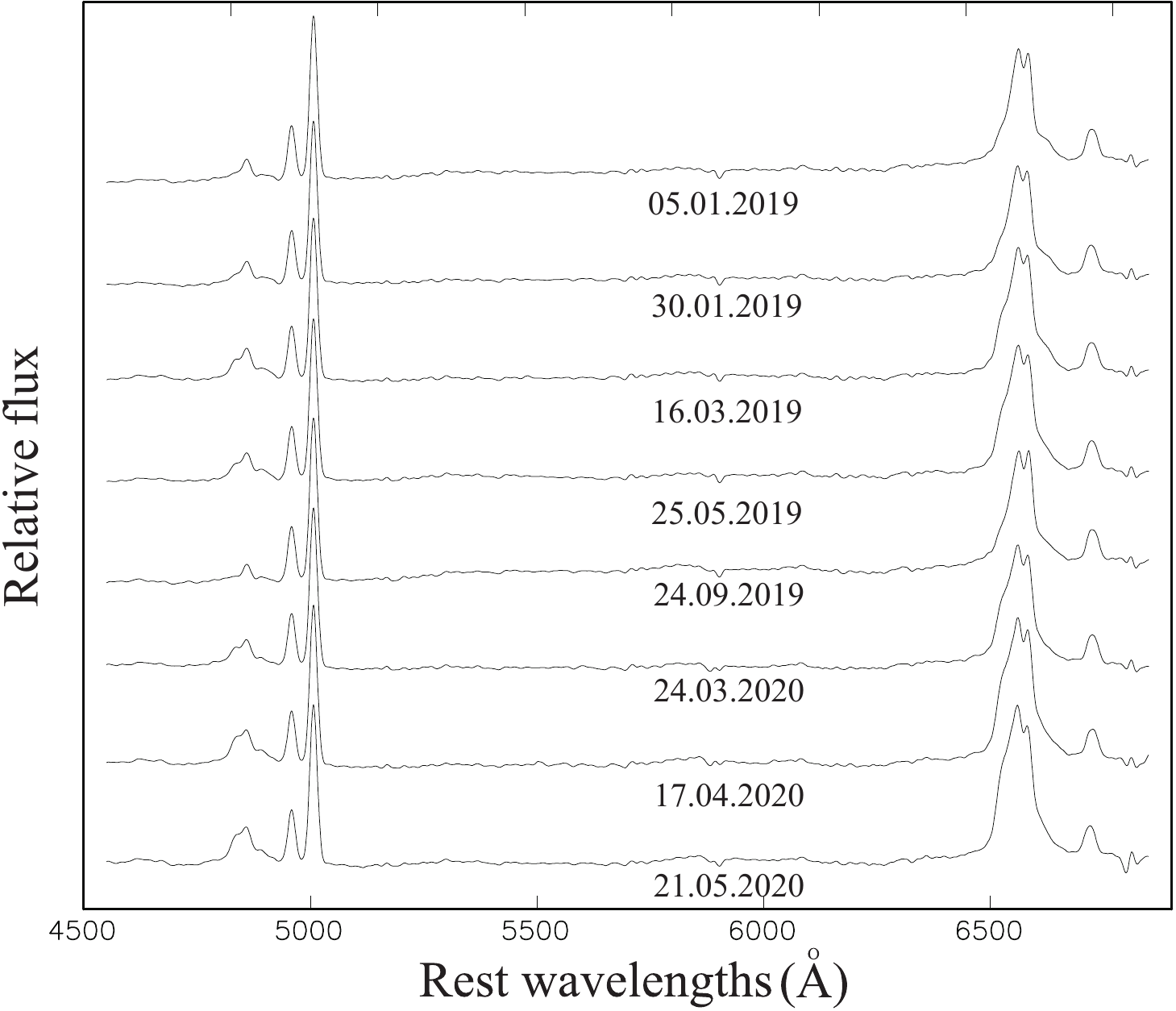}
    \caption{The selected host-galaxy subtracted spectra, obtained in the WIRO for those spectra covering the H$\alpha$ and H$\beta$ regions. All spectra are normalized to the [O III]$\lambda$5007 intensity and shifted for comparision.}
    \label{somewiro}
\end{figure*}

\begin{figure}
	\includegraphics[width=\columnwidth]{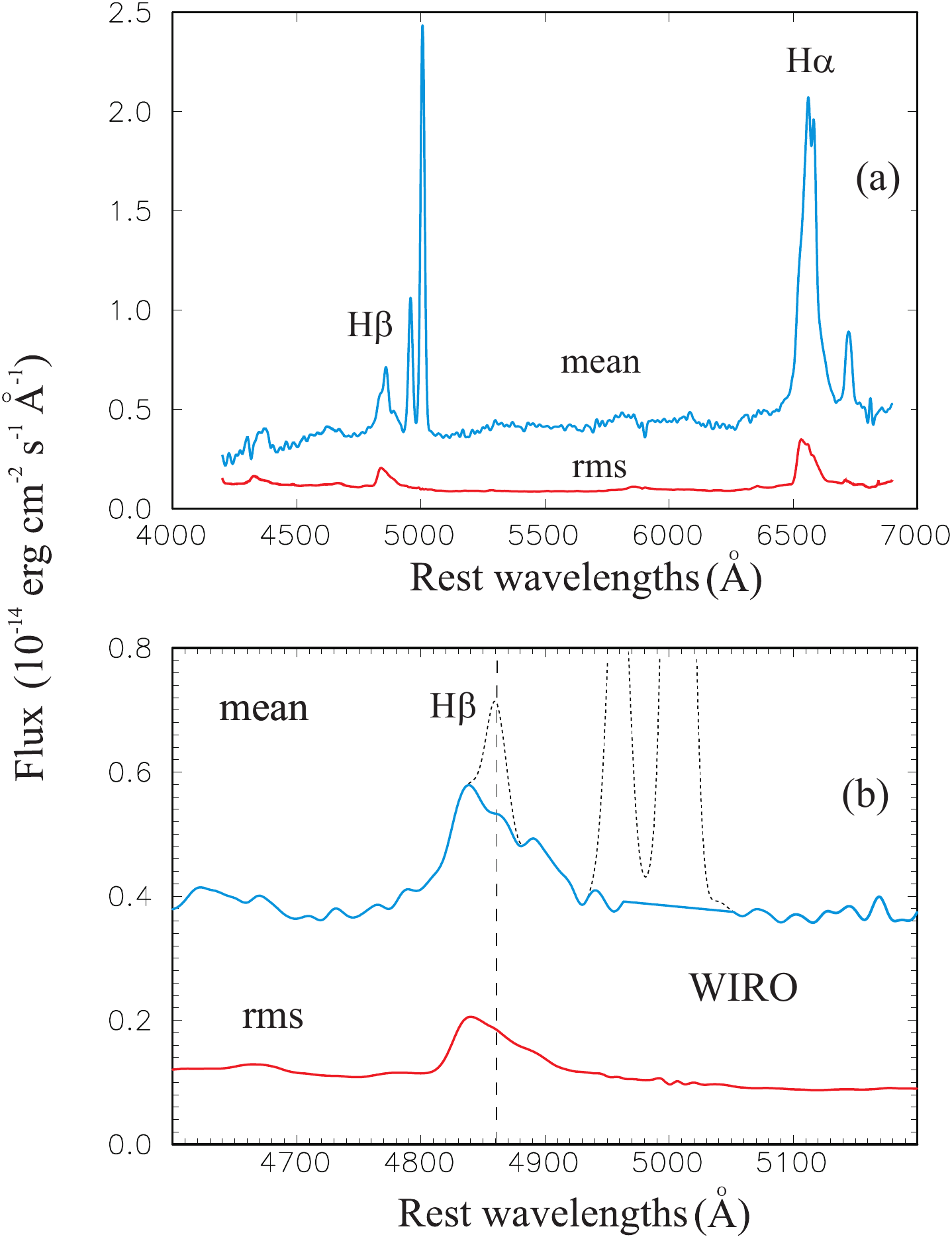}
  \caption{Mean and rms spectra (WIRO) in the rest-frame for the 
full spectral region (a) and just for the H$\beta$ region (b). In the bottom 
panel the dotted lines are the narrow H$\beta$ and [OIII]$\lambda$4959,5007 in the 
mean spectrum (see more details in the text), while the dashed line 
indicates 4861 \AA~ in order to illustrate the asymmetric H$\beta$ profile more 
clearly.
        }    
    
   \label{fig5}
\end{figure}

\begin{figure}
	\includegraphics[width=\columnwidth]{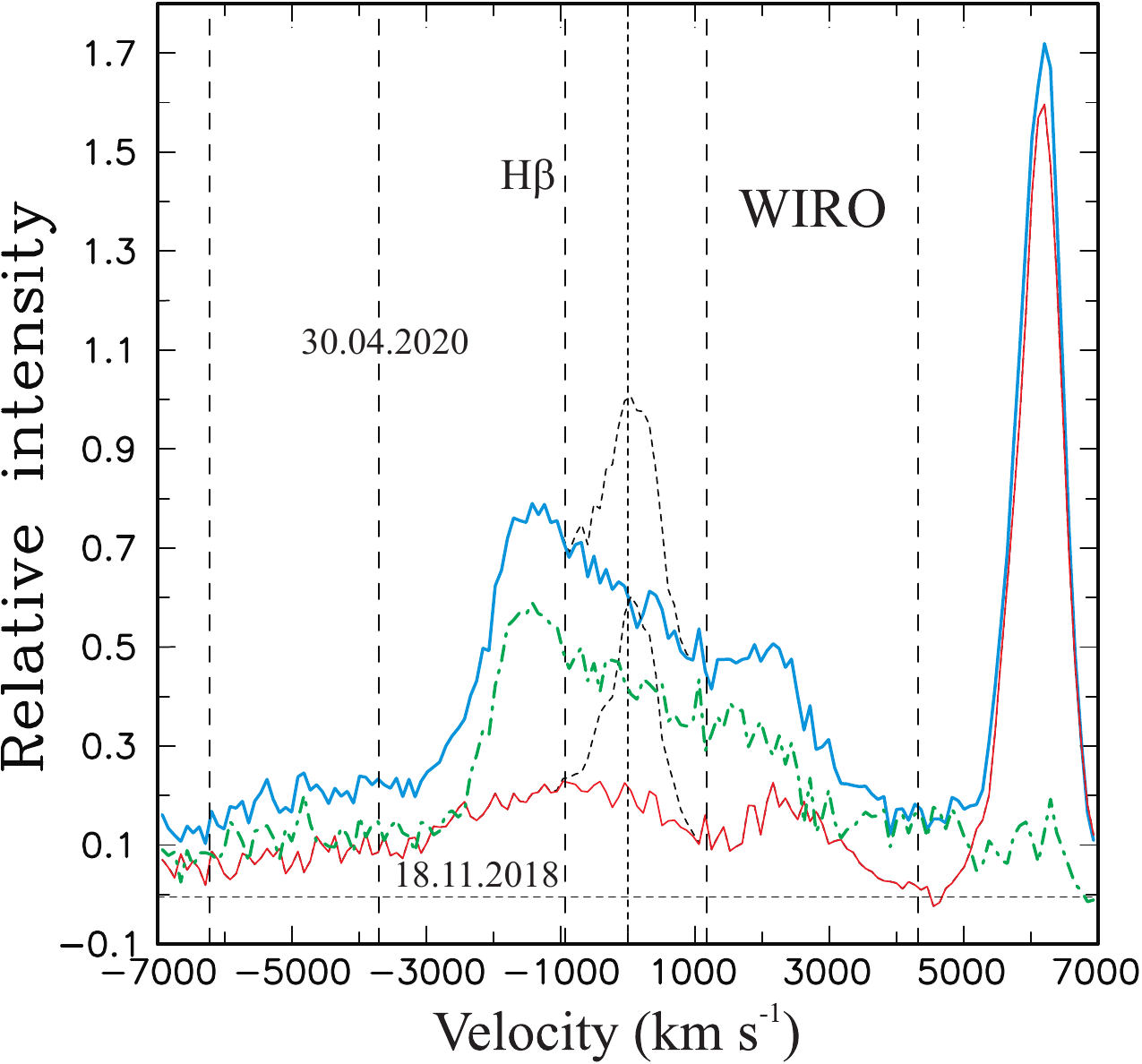}
    \caption{Profiles of broad H$\beta$ for the maximal (30.04.2020) and  minimal (18.11.2018) states (with the narrow components removed) showing intrinsic variations of the line. The difference between the maximal and  minimal states is shown by the dash-dotted line. The vertical dashed lines (1-5) mark the regions for integration of I(H$\beta$)~(1-2), I(H$\beta$(Blue))~(3-4) and I(H$\beta$(Red))~(5-2). See text for details.}
    \label{fig6}
\end{figure}

\begin{figure*}
	\includegraphics[scale=0.9,angle=0,trim=0 0 0 0]{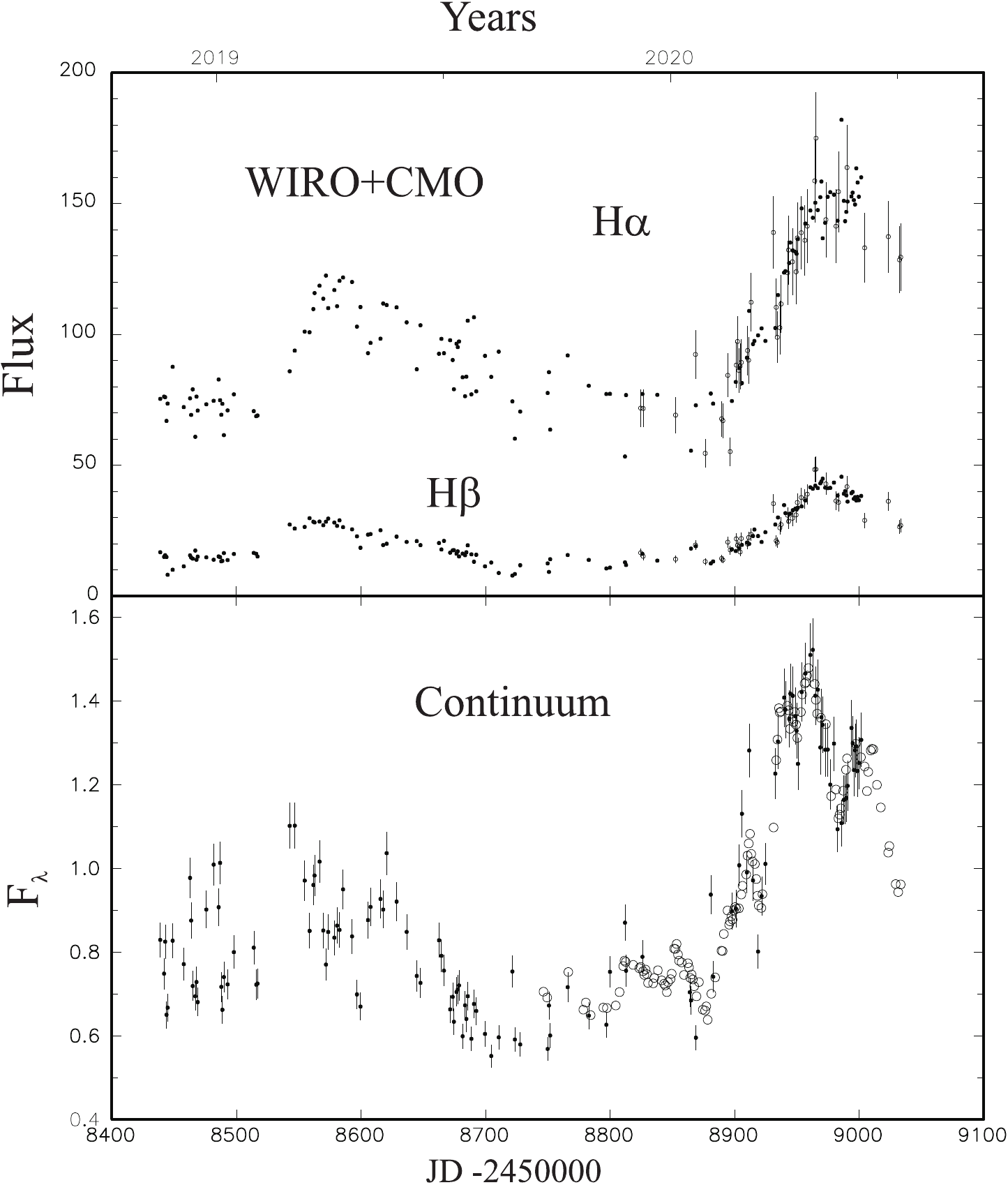}
    \caption{Combined H$\alpha$, H$\beta$ and continuum (WIRO and CMO) light curves given in fluxes (in units 10$^{-14}$erg s$^{-1}$cm$^{-2}$ for lines and 10$^{-14}$erg s$^{-1}$cm$^{-2}$\AA$^{-1}$ for contimuum). The CMO data are shown as open circles, the WIRO data as dots. The errors for CMO continuum (from photometry) are not larger than the size of circles. The error bars are shown just for the WIRO continuum values and the CMO H$\alpha$ and  H$\beta$ data.  The errors for WIRO spectral data are not shown. See text for details.}
    \label{fig7}
\end{figure*}

We obtained optical spectra (range 3500-7000 \AA) using the 2.5-m telescope of the Caucasian Mountain observatory (CMO) of SAI MSU (further CMO, see Paper~I for details) during 40 nights from 7 Dec. 2019 to 4 Jul. 2020 with a cadence of about 5 days. The telescope is equipped with the medium resolution optical Transient Double-beam Spectrograph (TDS), with Andor Newton 940P cameras using CCDs E2V CCD42-10, and a volume phase holographic grating \citep{Potanin2020}.  The slit width in most cases was 1$\farcs$0, and 1$\farcs$5 for 9 Dec. 2019.  The exposure times were from 1200~s to 3600~s, giving a spectral resolution and S/N ratio of $\sim$4~\AA \ and 50-120 near H$\beta$, and $\sim$3~\AA \, and 70-140 near H$\alpha$. Standard data reduction procedures were performed using self-developed Python3 scripts (see Paper~I for details). The spectrum of the nuclear region representing the flux collected within an aperture of 6 arcsec centered on the nucleus was flux-calibrated using spectrophotometric standard stars. We corrected the spectra for the host galaxy contamination which was estimated by scaling the off–nucleus spectrum measured between 6 and 16 arcsec on both sides from the centre of the nucleus. The narrow emission lines of NGC~3516 extend more than 20 arcsec from the nucleus and are clearly seen in our spectra \citep[see e.g.,][]{wa93}. We linearly interpolated under these extended narrow lines to remove them from the host galaxy spectrum. The resulting host galaxy spectrum was then scaled by a factor that was adjusted until the stellar absorption features matched those in the object spectrum.  To scale H$\beta$ we used an assumption that the [\ion{O}{iii}]$\lambda$5007 line was not variable during the months of our monitoring. For  extended sources (as in the case of NGC3516) this method might be not quite suitable and it is needed to correct the data some way for seeing effects  \cite[e.g.,][]{wa93, Feng2020}. Additionally (as a part of our standard calibration procedure), we corrected the data for seeing effects using our optical photometry, which was obtained on the same dates as the spectra. We assumed that the optical continuum flux variability at different wavelengths is synchronous with a linear dependence between those wavelengths.  This assumption is supported by our cross-correlation and RM investigations for the variability in the $U$ and $B$ filters (see further Sec. 3.3 and 
Table~\ref{Tab1d}).  We used the $U$ band photometry to correct our spectra since in this band the amplitude variation is a few times larger than in the $B$ and $V$ bands but the accuracy is about the same for these bands. 
We estimate the accuracy of such the calibration of the host-corrected spectra to be $\sim$1\%. So, the errors of the emission-line intensities are mostly dependent upon the continuum level fitting errors. 
As an example, Fig.~\ref{fig3} shows the full spectrum for 24 Apr. 2020 (both observed and corrected for the host galaxy input). Fig.~\ref{somecmo} shows some spectra (corrected for the host galaxy)  for selected dates during January -- June 2020. Dramatic variations of the broad lines and of the UV continuum are visible.
Fig.\ref{rmscmo} shows the mean and root-mean-square (rms) spectra in the H$\beta$ 
region for CMO data  (for details see \cite{Du2018}). We note that the 
variation of seeing and small errors in wavelength calibration will 
affect the shapes of narrow lines \citep{Peterson2004, Feng2020}. Meanwhile, the [O III] (as well as other forbidden lines) and 
the narrow H$\beta$ emission lines which are seen in the rms spectra are 
very small relative to these lines in the mean spectrum and that is an 
indication of reasonable calibration procedure. As seen from the rms 
spectra, the variability in broad H$\beta$ as well as in H$\alpha$ and H$\gamma$  lines was 
significant and asymmetric.

In Fig.~\ref{fig2} we present the variations of the broad H$\alpha$ and H$\beta$ flux (here and further, see the wavelength ranges for integrating the line intensities   in Table~\ref{Tab2w}) relative to the [\ion{O}{iii}]$\lambda$5007 flux. 
The same ratios were calculated for the blue wing of H$\beta$  -- I(H$\beta$(Blue)) and for the red wing of H$\beta$  -- I(H$\beta$(Red)). 
Previous to integration, we removed the narrow components with zero velocity in H$\alpha$ and H$\beta$,  by fitting them 
with  a Gaussian plus a linear background.  The latter may represent some contribution of underlying broad components, which can be nearly linear over a small wavelength range. Similarly, the [\ion{N}{ii}] $\lambda$6548,$\lambda$6584 narrow lines were removed from H$\alpha$. We also integrated fluxes of H$\gamma$ (blended with [\ion {O}{iii}]$\lambda$4363) and  \ion{He}{ii} $\lambda$4686 (blended with \ion{Fe}{ii}) relative to [\ion{O}{iii}]$\lambda$5007  without removing narrow components (further, for brevity, H$\gamma$ and \ion{He}{ii}, respectively).   
The variations of \ion{He}{ii} are shown in Fig.~\ref{fig2}. We present  all these values and associated errors in  Table~\ref{Tab2d}. The errors were estimated from the internal consistency of the data for close dates.
  Fig.~\ref{fig4} shows spectra for the H$\beta$ region for dates of minimum and maximum flux 
(10 Feb. 2020 and 24 Apr. 2020), as well as the difference spectrum, 
and the regions of integration for I(H$\beta$), I(H$\beta$(Blue)) and I(H$\beta$(Red)).

\begin{table}

\centering
\caption{Spectral lines and the wavelength ranges (in \AA) for integrating the line intensities.}

\begin{tabular}{ccc} \hline
   Spectral line& wavelengths (CMO) & wavelengths (WIRO)\\
   \hline
   H$\alpha$ & 6380 -- 6700 & 6480 -- 6670 \\
   H$\beta$&4757 -- 4936 &   4760 -- 4930 \\
   H$\beta$(Blue)& 4815 -- 4853 & 4800 -- 4845 \\
   H$\beta$(Red)& 4876 -- 4936 & 4880 -- 4930 \\
   H$\gamma$& 4260 -- 4406 &   --  \\
   He II & 4544 -- 4695 &  --  \\ \hline

 \\

\end{tabular}

\label{Tab2w}

\end{table}

\begin{table*}
\centering
\caption{Results of optical spectroscopy with the 2.5-m telescope at CMO. ( Note:  The full version of this table is available in its entirety in machine-readable form.)}

\begin{tabular}{ccccccc} \hline
   J.D.-2450000 & H$\alpha^*$   & H$\beta^*$ & H$\beta$(blue)$^*$ & H$\beta$(red)$^*$& H$\gamma^*$&  He II$^*$ \\
   \hline
8824.57&  3.26 $ \pm$  0.33 &  0.75 $ \pm$  0.08 &  0.24 $ \pm$  0.02 &  0.22 $ \pm$  0.02 &  0.40 $ \pm$ 0.05 &  0.16 $ \pm$  0.03\\
8826.59&  3.26 $\pm$  0.33 &  0.69 $ \pm$  0.07 &  0.23 $ \pm$  0.02 &  0.22 $ \pm$  0.02 &  0.39 $ \pm$ 0.05 &  0.18 $\pm$  0.04\\
8852.59&  3.14 $ \pm$  0.31 &  0.64 $ \pm$  0.06 &  0.23 $ \pm$  0.02 &  0.16 $ \pm$  0.02 &  0.35 $ \pm$ 0.05 &  0.15 $ \pm$  0.03\\
8868.57&  4.20 $ \pm$  0.42 &  0.89 $ \pm$  0.09 &  0.34 $ \pm$  0.03 &  0.20 $ \pm$  0.02 &  0.62 $ \pm$ 0.08 &  0.24 $ \pm$  0.05\\
8876.40&  2.48 $ \pm$  0.25 &  0.60 $ \pm$  0.06 &  0.22 $ \pm$  0.02 &  0.15 $ \pm$  0.02 &  0.36 $ \pm$ 0.05 &  0.20 $ \pm$  0.04\\
... & ... & ... & ... & ...& ... & ...
\\ \hline 

 \\

\end{tabular}

 \begin{tablenotes}

	    \item $^{*}$ Flux value relative to  [\ion{O}{iii}]$\lambda$5007 which mean-weighed  flux is 2.2 $\times$10$^{-13}$ erg cm$^{-2}$ s$^{-1}$.

    \end{tablenotes}

 \label{Tab2d}

\end{table*}

\subsection{Spectral observations at WIRO}

A significant part of our spectral data set was obtained with the 2.3 meter Wyoming Infrared Observatory (WIRO) telescope from Nov. 2018 until 31 May 2020 with a cadence of about 4 days but which includes a seasonal gap. The observation and reduction methods were the same as described by \cite{Du2018}. The primary differences compared to the CMO spectral data  consist of a wider aperture (5 arcsec oriented north-south) and correspondingly lower spectral resolution ($\sim$10 $\AA$ for a uniformly filled slit).  Typically we observed NGC~3516 in three 300 second exposures and flux calibrated the spectra using spectrophotometic standards G191B2B, Feige 34, HZ 44, or BD+28$^{\circ}$4211. We corrected the spectra for the host galaxy contamination in the same way as it was described before for the CMO spectral data. Fig.~\ref{somewiro} shows some spectra (corrected for the host galaxy)  for selected dates during January – June 2020. Dramatic brightenings of  the broad H$\beta$ line occurred twice.
   The  mean and root-mean-square (rms)  spectra in H$\beta$ region (just WIRO data) for NGC~3516 are shown in Fig.\ref{fig5} (for details see \cite{Du2018}). The [O III] and the narrow H$\beta$ emission lines in the rms spectra are not seen, since there is no variability and that is an indication of correct calibration procedure.  As it is seen from the rms spectra the variability in the broad H$\beta$ and H$\alpha$ lines was significant and asymmetric.

The  spectral data  obtained at WIRO and CMO were processed completely independently, and therefore, as well as due to different spectral resolutions, the spectral intervals for integration within the broad H$\beta$ line were set slightly differently.
The  H$\alpha$ and H$\beta$ fluxes see Table~\ref{Tab2w} normalized to the  \ion{O}[{III}]$\lambda$5007  flux.  The same ratios but only for the blue H$\beta$(Blue) and red H$\beta$(Red) wings of the line were calculated.. Narrow emission-lines were removed before the integration in the same way as it was done with the CMO spectral data.  

Fig.~\ref{fig6} shows spectra of the H$\beta$ region for minimum and maximum fluxes 
(18 Nov. 2018 and 30 Apr. 2020) for our WIRO observations, as well as 
the difference spectrum, and the areas of integration for I(H$\beta$), I(H$\beta$(Blue)) and I(H$\beta$(Red)). 

The continuum flux in the WIRO data for each spectrum was estimated as 
the median value in the region 5075 -- 5125\AA. We have found a high 
correlation coefficient between the CMO and the WIRO data (r$>$0.9) for 
the H$\alpha$ and H$\beta$ line variations. Since observations were carried out 
using instruments of different apertures, the [\ion{O}{iii}]$\lambda$5007 flux was 
about two times larger in the WIRO data than in the CMO.
Fig.~\ref{fig7} shows combined light curves of the WIRO and CMO data for H$\alpha$, H$\beta$ and the continuum reduced to the same system as that of the CMO spectral data and also fluxes based on photometry I($U$) (using a linear least squares regression). The WIRO spectral data are  presented in Table~\ref{Tab3d}.

\begin{table*}
\centering
\caption{Results of optical spectroscopy with the 2.3-m  WIRO telescope. ( Note:  The full version of this table is available in its entirety in machine-readable form.)}
\begin{tabular}{cccccc} \hline
   J.D.-2450000 & H$\alpha^*$   & H$\beta^*$ & H$\beta$(Blue)$^*$ & H$\beta$(Red)$^*$ &Cont5100$^a$  \\
   \hline
8438.90&   3.42 $ \pm$  0.34 &  0.76 $ \pm$  0.04 &  0.20 $ \pm$  0.02 &  0.27 $ \pm$  0.03 &  1.62 $ \pm$  0.08 \\
8442.01&   3.46 $ \pm$  0.35 &  0.68 $ \pm$  0.03 &  0.18 $ \pm$  0.02 &  0.22 $ \pm$  0.02 &  1.60 $ \pm$  0.08 \\
8442.94&   3.45 $ \pm$  0.35 &  0.71 $ \pm$  0.04 &  0.20 $ \pm$  0.02 &  0.24 $ \pm$  0.02 &  1.62 $ \pm$  0.08 \\
8443.98&   3.04 $ \pm$  0.30 &  0.68 $ \pm$  0.03 &  0.19 $ \pm$  0.02 &  0.22 $ \pm$  0.02 &  1.56 $ \pm$  0.08 \\
8444.88&   3.34 $ \pm$  0.33 &  0.37 $ \pm$  0.02 &  0.09 $ \pm$  0.01 &  0.13 $ \pm$  0.01 &  1.57 $ \pm$  0.08 \\
... & ... & ... & ...& ... & ...
\\ \hline
 \\    

\end{tabular}
 \begin{tablenotes}
        \item $^{*}$ Flux value reduced  to the CMO system  relative to  [\ion{O}{iii}]$\lambda$5007 which mean-weighed  flux is 2.2 $\times$ 10$^{-13}$ erg cm$^{-2}$ s$^{-1}$.
        \item$^a$ Flux in continuum at 5100~\AA~in $10^{-14} $erg cm$^{-2}$ s$^{-1}$ ~$\AA^{-1}$ units.
    \end{tablenotes}

 \label{Tab3d}
\end{table*}

\section{Light curves and time series analysis}
\subsection{Light curves}
As can be seen from the light curve since 1999 (Fig.~\ref{fig1}), NGC~3516 was on average significantly brighter in 1999-2012 than in 2013-2019. After a maximum in 2007 (high state), the brightness of the object decreased to a minimum in 2014 (when the CL event was discovered by \cite{sh19}). After that, the object was in the low state, but from the end of 2015 to the middle of 2016 some  re-brightening was observed. A new CL event corresponding to this high state was observed by \cite{Oknyansky2020b}. After this  episode, the object was very low again for about 3 years. The relative amplitude of the variability in the optical bands was small, largely due to the relatively large contribution from the host galaxy light within the aperture. The moderate optical variability during this period corresponded to a much more significant variability in the UV, which is explained in part by lower contamination in this wavelength region from the host galaxy.

The optical continuum and Balmer line light curves (see Fig.~\ref{fig7}) for 2018-2020 show two epochs of brightening: the first one, at the beginning of 2019, was relatively small, and the second one, which started after a deep minimum  at the end of January 2020 and continued till July 2020, was very strong. That strong brightening was accompanied by a significant enhancement of the broad emission lines and variability of their profiles (see Fig.~\ref{fig4},\ref{fig6}). During the spring of 2020, several fast flares were observed in all wavelengths (from optical to X-ray ) on a timescale of a few weeks.  As can be seen from  Fig.~\ref{fig2} the variations in $B$, $U$, and $UVW2$ in Feb.-Jun. 2020 took place synchronously without any obvious differences. The maxima of these flares were recorded at the all wavelengths from the optical to X-ray. The maxima reached in X-rays (on 2 Mar. and on 1 Apr.2020) were followed by the maxima in UV and optical bands some 1-2 days later. At the maximum (on 1 Apr. 2020) the X-ray flux reached the highest level over the entire history of observations of this object by Swift (nearly 22 times brighter than at the minimal observed level on 27 Jun. 2014).  After the maximum, the brightness of the object began to weaken in X-rays but continued to grow in UV/optical bands till the middle of Apr. 2020. So, we have detected uncorrelated changes in X-rays and the UV/optical continuum in Apr.- Jul. 2020.


 The visual analysis of the variations of broad H$\beta$ during this 7 month period (Dec. 2019--Jul. 2020) shows that the line follows the variations of the continuum but with a delay of about 2-3 weeks (see Fig.~\ref{fig2}). Such a strong increase in H$\beta$ observed on 22 Mar. corresponds to the maximum at the beginning of Mar. 2020. The strong maximum observed in the line at the end of April 2020 corresponds  to the maximum in the continuum at the beginning of April. The minimum of the line flux on 10-16 Feb. 2020 corresponds to the minimum of the continuum at the end of Jan. 2020. The local maximum of the line on 17 Jan. 2020 echoes the continuum maximum of 3 Jan. 2020.
These visual impressions are confirmed quantitatively by our RM results produced by three different methods that are in good agreement (see details below).

\subsection{Reverberation mapping methods}

To investigate possible lags between variations in the spectral lines and continuum, we used three different RM methods commonly used in the literature: ICCF (traditional interpolation cross-correlation techniques \citep{Gaskell1986,  Peterson1998, Peterson2004}), MCCF (which is a modification of ICCF \citep{Oknyansky1993}, and an alternative method of measuring reverberation time lags - JAVELIN (Bayesian analysis \citep{Zu2011,Zu2013}) -- formerly known as Stochastic Process Estimation for AGN Reverberation (SPEAR). All these methods have been used many times \citep[see e.g.,][]{Du2014,Grier2012, Li2019, Oknyansky2017a, Oknyansky2019a}. MCCF and ICCF are very similar; MCCF is a modernisation of ICCF to reduce the interpolation errors. In MCCF we used only those interpolated points that were separated in time from the nearest observation points by no more than a limit value MAX=5 days. For interpolation we used just one data set that had better accuracy and higher time-cadence. ICCF is more often used in publications and  we implemented a publicly available PyCCF code \citep{Sun2018}, adapted from the original ICCF code \citep{Peterson1998}. In both methods, ICCF and MCCF, the errors of time delays were estimated in the same way following \cite{Peterson2004}. JAVELIN is also very popular and we used it in the same way as it was discussed in our previous publications \citep[e.g.,][]{Oknyansky2017a, Oknyansky2019a}.

Using these techniques, the lag between X(t) and Y(t) can be estimated in at least two ways: from ${\tau}_{peak}$ corresponding to maximum correlation in ICCF and MCCF or equivalently the number N in the histogram (JAVELIN), and from the centroid ${\tau}_{cent}$ \citep[see e.g.,][]{Oknyansky2017a}. The discussion and references on difference and advantages of `peak' and `centroid' estimations can be also found in \cite{, Peterson2004}. We prefer to use ${\tau}_{peak}$ when we have principal peaks in the cross-correlation functions well defined and isolated, since it depends on fewer free parameters. The results of our RM (MCCF and JAVELIN) are partially presented in Figs.~\ref{fig8}-\ref{fig11}, whereas all the results obtained with three different methods are collected in Table~\ref{Tab1}.

\subsection{Time delay measurements}

\begin{figure}
	\includegraphics[scale=0.9,angle=0,trim=0 0 0 0]{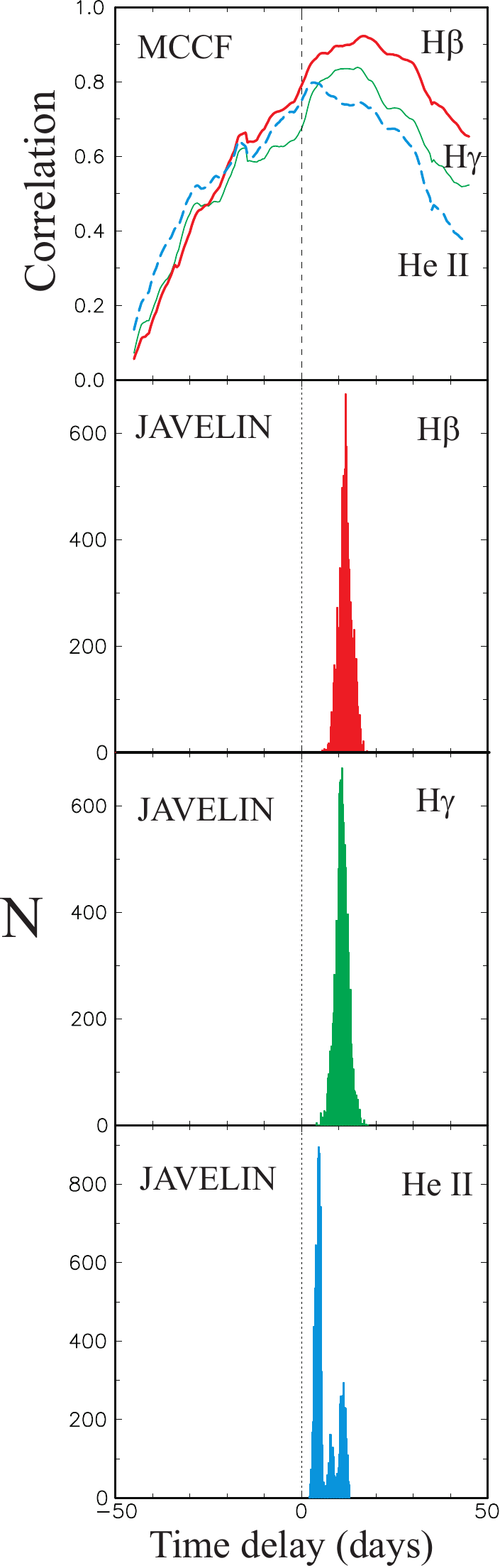}
    \caption{Top panels: RM by the MCCF method (just for the CMO data) for H${\beta}$ (solid line), H${\gamma}$ (thin line) and He II${\lambda}$4686 (dashed line) relative to optical continuum I($U$). Bottom panels:  RM by JAVELIN using the same data. The histograms are shown for each of these lines independently. The vertical dashed lines indicate zero lag.}     \label{fig8}
\end{figure}

The emission-line time lags multiplied by the speed of light ($c$) provide the emissivity-weighted size scales of the regions corresponding to the variable broad lines. In case of time lags associated with the broadband X-ray/UV/optical continuum responses, they may correspond to an effective radius of an accretion disk (although there may be a component of continuum from the broad line region (BLR) as well \citep[see e.g.,][]{Korista2001}.

 The RM for continuum flux variations in the optical ($U$) and $UVW2$ bands confirms the results of visual inspection -- these variations show a very high correlation with a maximum of $r\sim0.97$ near zero lag. However a small possible time shift ($\sim$ 0.5 days) between these variations was found by MCCF and JAVELIN  (see Table~\ref{Tab1}), although the confidence of this result is not high due to relatively large gaps in the $Swift$ observations. The X-ray flux variations correlate well with the optical ones in the interval 19 Feb.- 10 Apr. 2020 ($r\sim0.87$) with a delay of about 1-2 days, but the correlation drops if all the X-ray data through the end of June 2020 are included. Unfortunately, there are not enough X-ray data to provide a high level of confidence for this result.
 The variation of the X-ray flux was ahead of the I($UVW2$) variations by $\sim$ 1.6 days  during  8 Apr. 2012- 10 Apr. 2020. That is in agreement with the result published by  \cite{bu17} who used in part the same $Swift$ data. That is also in agreement with another published estimate \citep{nd16} of the delay of about 2 days between the optical and X-ray  variations.

Our RM analysis was done independently for the CMO and WIRO spectral data with our CMO photometry, as well as for the combined light curves using our three different methods. We investigated time delays relative to the continuum variation not only for different broad emission lines (Balmer lines, He II), but also for the blue and red wings of H${\beta}$ independently.

\begin{figure}
	\includegraphics[width=\columnwidth]{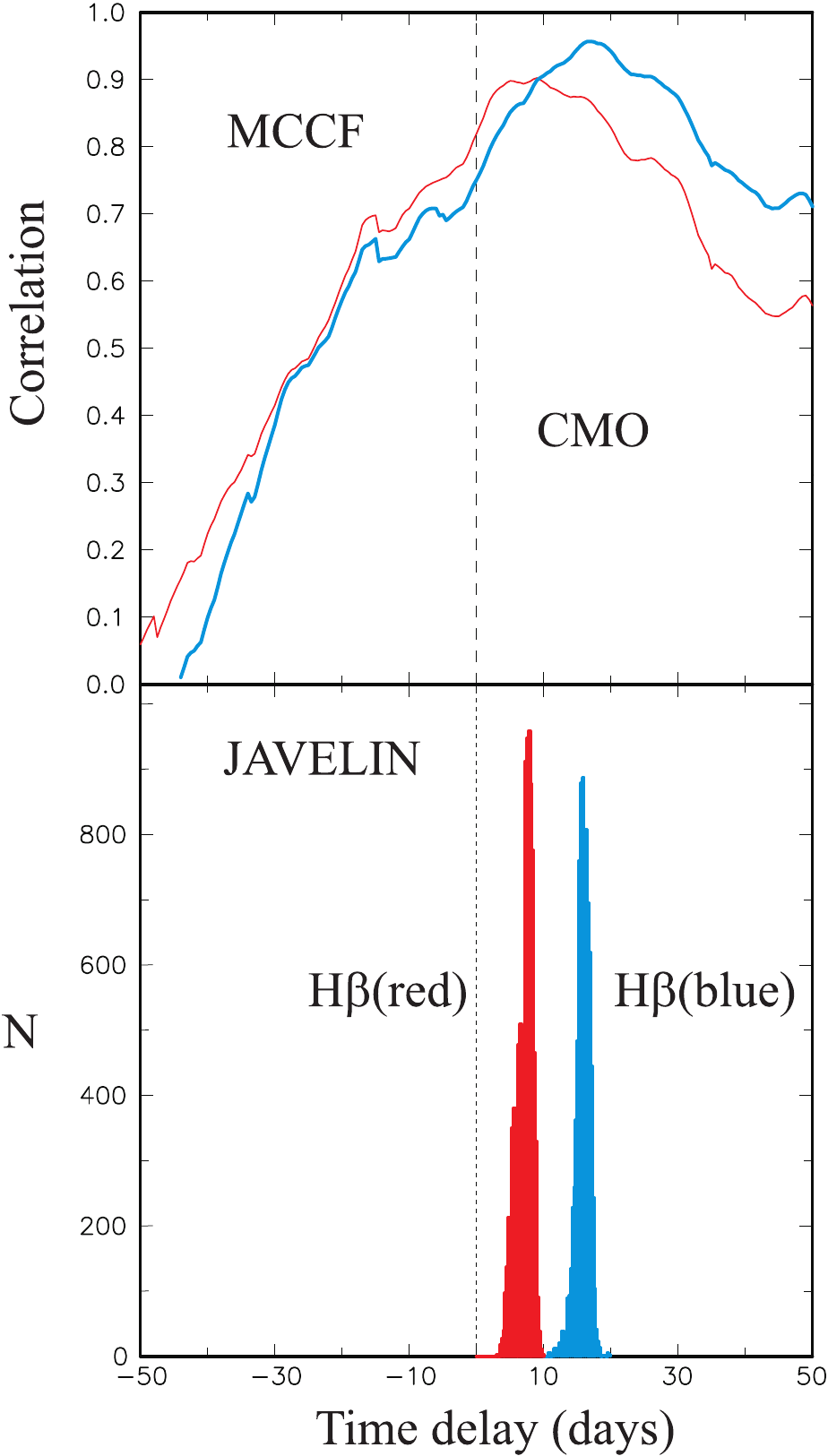}
    \caption{Top panels: RM by the MCCF method  (just for CMO data)  for H${\beta}$(blue) (solid line), H${\beta}$(red) (thin line) relative to optical continuum I($U$). Bottom panel:  the RM histogram by the JAVELIN using the same data.  The vertical dashed lines indicate zero lag.}
   \label{fig9}
\end{figure}

\begin{figure}
	\includegraphics[width=\columnwidth]{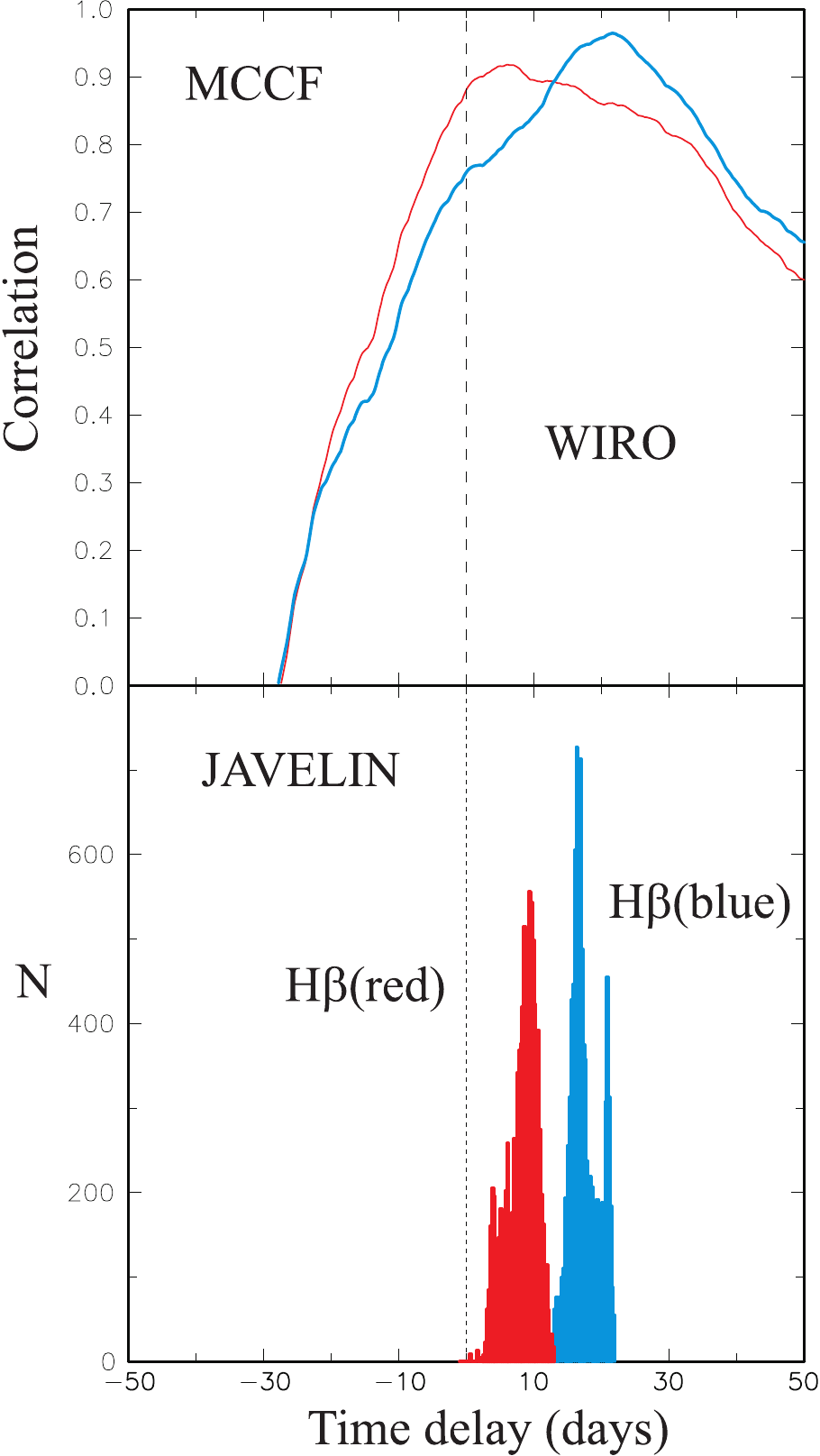}
    \caption{Top panels: RM by the MCCF method  (just for the WIRO spectral data)  for H${\beta}$(blue) (solid line), H${\beta}$(red) (thin line) relative to optical continuum I($U$) from the CMO photometry. Bottom panel:  the RM histogram by JAVELIN using the same data.  The vertical dashed lines indicate zero lag.}     \label{fig10}
\end{figure}

\begin{figure}
	\includegraphics[scale=0.9,angle=0,trim=0 0 0 0]{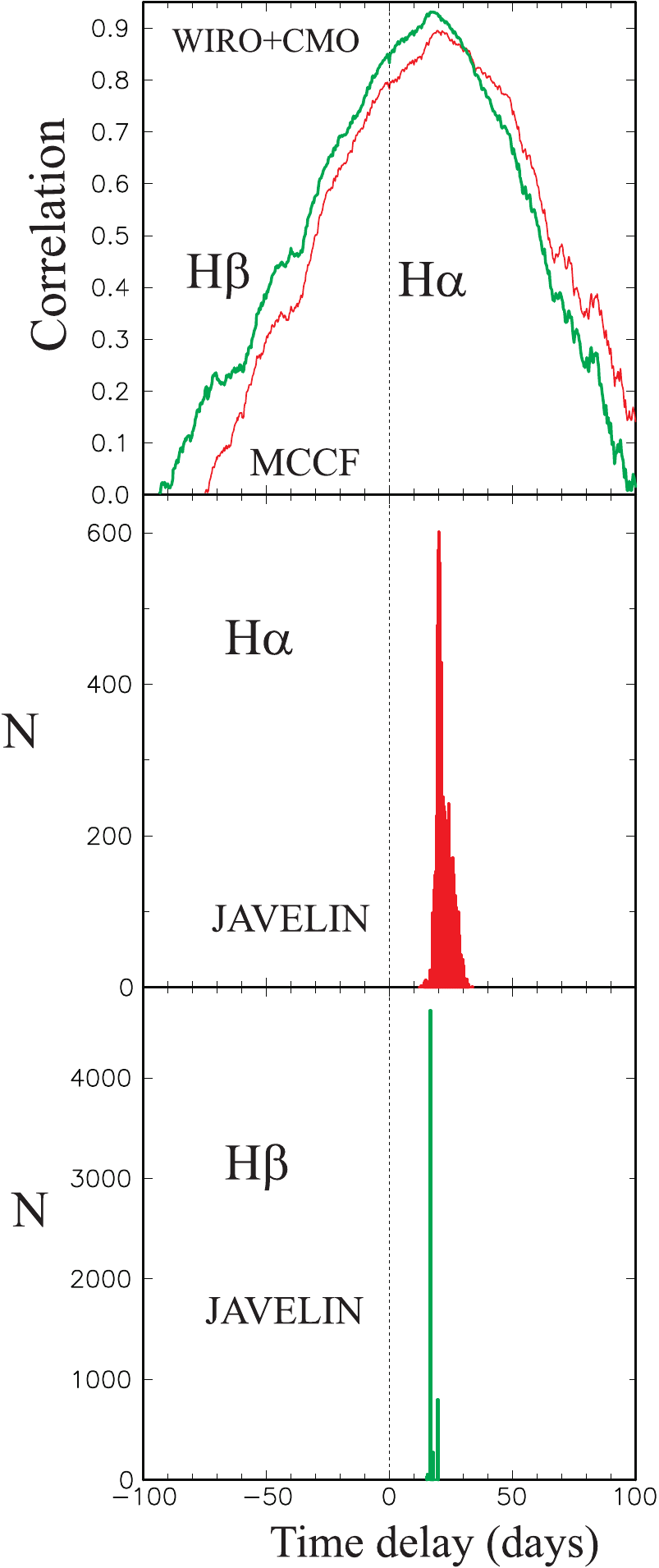}
    \caption{Top panels: RM by the MCCF method  (for the combined CMO and WIRO data) for H${\beta}$ (solid line), H${\alpha}$ (thin line) relative to optical continuum (see text). The RM histograms by JAVELIN (using the same data) are shown for  H${\alpha}$ at the middle panel and for H${\beta}$ at the bottom panel. The vertical dashed lines indicate zero lag.}     \label{fig11}
\end{figure}

A quick inspection of Figs.~\ref{fig8}-\ref{fig11} and Table~1 confirms that the results obtained with these different methods correspond well with one another. MCCF and ICCF give very similar results for the case when we have well sampled light curves. Some possible differences in the lags obtained with JAVELIN and ICCF have been discussed many times \citep[see e.g.,][]{Lira2015}. For the rest of this section, we will refer in the text only to ${\tau}_{peak}$ obtained with the MCCF, but all the results can be found in Table~\ref{Tab1}.

Fig.~\ref{fig8} demonstrates the MCCF (top panel) and JAVELIN results (bottom panels) for H${\beta}$, H${\gamma}$ and He II${\lambda}$4686 relative to optical continuum I($U$) using just the CMO data. The time delays are well defined for these RM results. The H${\beta}$ variations lag by about 17 days (see Table~\ref{Tab2d} for details and error values). H${\gamma}$  has a smaller delay of about 15 days, although the difference is not very significant. He II has the smallest delay of about 3 days. These relative time delays are seen independently in the MCCF, ICCF and JAVELIN results.

RM for H${\beta}$(Blue) and H${\beta}$(Red) relative to optical continuum I($U$) is presented in Fig.~\ref{fig9} (just for the CMO data) and Fig.~\ref{fig10} (the WIRO spectral data and the CMO continuum data). Both pictures provide consistent evidence that the blue wing's delay is about two times longer than that of the red one (about 17 and 9 days, respectively). We note that the CMO photometry started in Sep. 2019, but spectral monitoring began in Dec. 2019. The WIRO monitoring started in Nov. 2018 and finished at the end of May 2020. So, RM for the CMO and WIRO data corresponds for slightly different time intervals (Dec. 2019--Jul. 2020 and Sep.2019-May 2020, respectively).

\begin{figure}
	\includegraphics[width=\columnwidth]{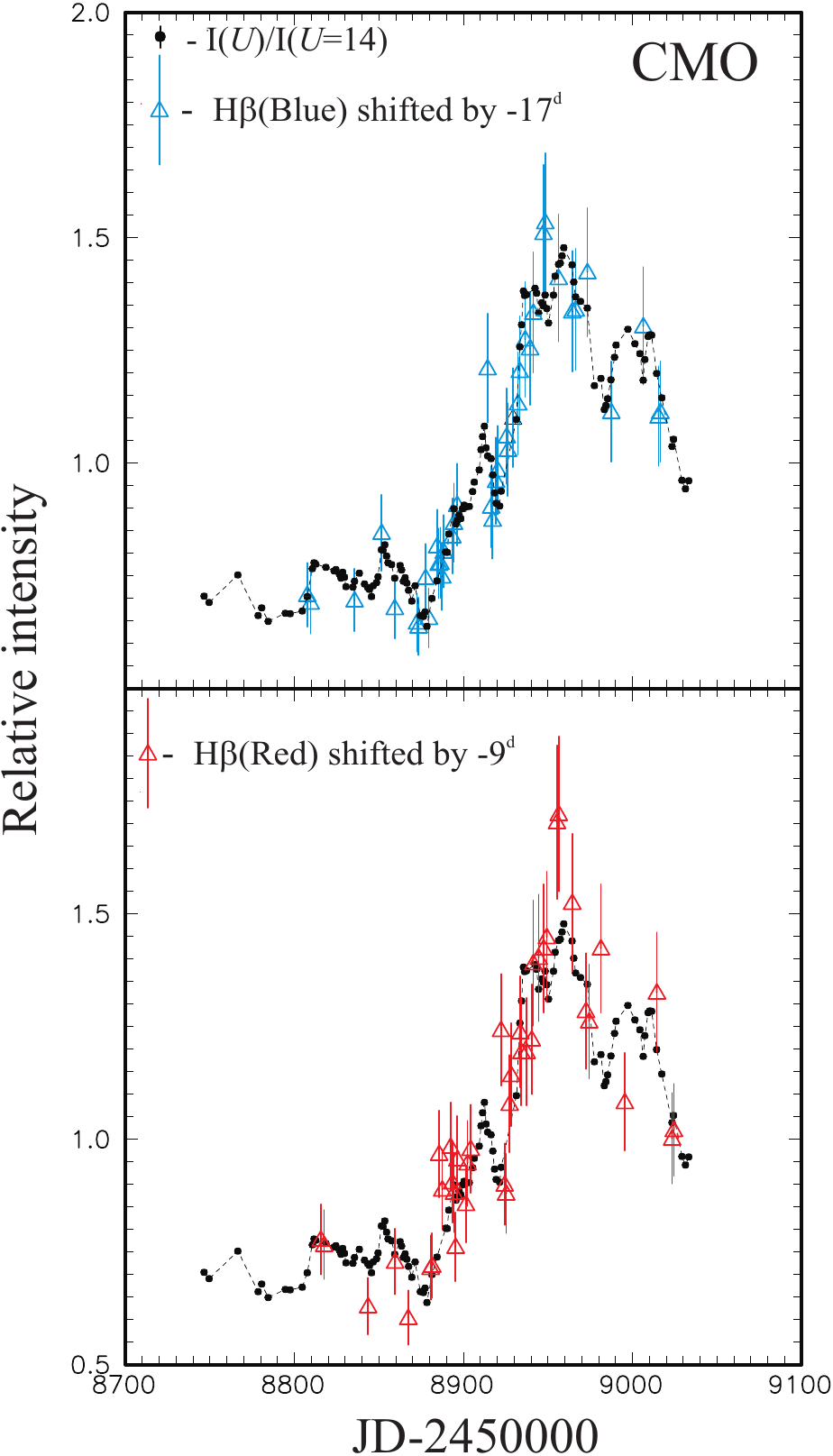}
    \caption{Top panel: Light curves for relative intensity  I$(U)$  (Intensity for U=14 mag corresponds to 1) and I(H{$\beta$}(Blue)) reduced to the I$(U)$ scale by the linear regression and shifted by -17 days. Bottom pane: The same as at the tope panel light curves   I$(U)$  and I(H{$\beta$}(Red)) reduced to the I$(U)$ scale by the linear regression and shifted by -9 days.   }     \label{fig12}
\end{figure}

RM for combined H${\alpha}$ and H${\beta}$ data relative to combined optical continuum I($U$) is presented in Fig.~\ref{fig11}. The maximum correlation coefficients are very high. H${\alpha}$ probably has a longer delay than H${\beta}$ (about 20 and 17 days, respectively). In Fig.~\ref{fig12} we show the light curve for H$\beta$(Blue) and  H$\beta$(Red)(just the CMO data) shifted  by $-$17 and $-$9 days and reduced by linear regression to the scale of I($U$). The correlation between the continuum and the blue wing of the line with a time delay of about 17 days is clearly seen from the plot.

\begin{table*}
\centering
\caption{Results from the RM analysis.  Lags for $X$ data sets are measured with respect to the Y data and are expressed in days.  A positive lag means the Y set leads the X
  variability. For MCCF the correlation values  in the peak $R_{max}$ are given. For interpolated MCCF and ICCF  $1\sigma$ confidence limits and centroids for interpolated cross-correlations  are presented. For JAVELIN  confidence limits are measured at a 95\%\ confidence level. Our preferred $\tau_{peak}$ values obtained by MCCF are given in bold. All delays are reduced to the rest system.}

\begin{tabular}{ccccccccc} \hline
  X &  Y        & J.D.-2450000  & Date            & MCCF   & ICCF& JAVELIN \\
  &&&& $R_{max}~~~~~~~\tau_{peak}~~ \tau_{cent}$&$\tau_{peak}~~ \tau_{cent}$&$\tau_{peak}$\\
  \hline \smallskip

 I($U$) (CMO) & I($B$) (CMO)&8746--9033& Sep.2019--Jul.2020 &0.998~~~${\bf0.4^{+0.2}_{-0.1}}~~~ 0.4^{+0.3}_{-0.3}$ & $ 0.4^{+0.2}_{-0.1}~~      0.2^{+0.5}_{-0.6} $ & $ 0.2^{+0.2}_{-0.1}$    \\ \smallskip

I($U$) (CMO) & I($UVW2$)&8899--9024& Feb.2020--Jun.2020 &0.967~~~${\bf0.4^{+0.3}_{-0.1}}~~~ 0.4^{+0.3}_{-0.3}$ & $ 0.4^{+0.1}_{-0.4}~~      0.0^{+0.1}_{-0.4} $ & $ 0.65^{+0.25}_{-0.25}$    \\ \smallskip

 I($UVW2$)&I(X-ray)& 6026--8943& Apr.2012--Apr.2020 &0.939~~~${\bf1.6^{+0.8}_{-0.6}}~~~ 1.4^{+0.9}_{-0.7}$ & $ 1.3^{+1.6}_{-1.5}~~      1.5^{+1.7}_{-1.8} $ & $ 1.35^{+0.25}_{-0.25}$    \\ \smallskip

H${\alpha}$ (CMO) &  I($U$) (CMO) &8824-9033& Dec.2019--Jul.2020 & 0.919~~${\bf16.9^{+0.6}_{-0.5}}~~ 16.6^{+0.7}_{-0.5}$ & $18.4^{+4.5}_{-3.8}~~18.6^{+3.9}_{-3.6}$&   $17.1^{+1.1}_{-4.4} $                  \\ \smallskip

H${\alpha}$ (WIRO)&  I($U$) (CMO) &8749-9001& Sep.2019--May 2020 &0.920~~${\bf20.9^{+6.0}_{-7.0}}~~ 18.8^{+6.4}_{-6.0}$ & $18.9^{+6.6}_{-6.2}~~19.7^{+4.1}_{-4.0} $ &    $19.1^{+2.8}_{-4.4}$                  \\ \smallskip

H${\beta}$ (CMO) &  I($U$) (CMO) &8824-9033& Dec.2019--Jul.2020 & 0.958~~${\bf16.7^{+3.5}_{-4.0}}~~ 15.9^{-3.4}_{+4.4}$ &$14.9^{-1.9}_{+4.6}~~ 15.0^{-4.1}_{+4.2}$&    $11.8^{+3.1}_{-3.1}$                  \\ \smallskip

H${\beta}$ (WIRO) &  I($U$) (CMO) &8749-9001& Sep.2019--May 2020 & 0.953~~${\bf15.3^{+2.2}_{-2.0}}~~ 14.0^{+2.4}_{-3.4}$ &$18.2^{+1.8}_{-6.2}~~ 15.9^{+2.2}_{-2.7}$ & $14.9^{+1.1}_{-0.1}$                    \\ \smallskip

H${\gamma}$ (CMO) &  I($U$) (CMO) &8824-9033& Dec.2019--Jul.2020 &0.839~~${\bf14.9^{+2.0}_{-5.0}}~~ 12.2^{+2.4}_{+5.4}$ &$12.5^{+5.8}_{-1.3}~~ 14.2^{+8.5}_{-4.6}$ &    $11.2^{+3.0}_{-3.0}$                 \\ \smallskip

He II (CMO) &  I($U$) (CMO) &8824-9033& Dec.2019--Jul.2020 & 0.798~~~${\bf3.2^{+2.5}_{-2.5}}~~~ 0.36^{-4.0}_{+5.0}$ &$3.7^{+6.8}_{-10}~~ 5.7^{+7.0}_{-9.7}$ &   $4.5^{+1.0}_{-2.0}$                   \\ \smallskip

H${\beta}$(Blue) (CMO) &  I($U$) (CMO) &8824-9033& Dec.2019--Jul.2020 &0.956~~${\bf16.9^{+2.0}_{-1.5}}~~ 15.8^{+2.4}_{-2.0}$ & $16.8^{+4.0}_{-0.8}~~ 19.3^{+5.3}_{-4.1}$&    $15.7^{+2.1}_{-2.0}$                   \\ \smallskip

H${\beta}$(Blue) (WIRO) &  I($U$) (CMO) &8749-9001& Sep.2019--May 2020 &0.965~~${\bf21.4^{+2.0}_{-2.4}}~~ 19.1^{+2.4}_{-2.4}$ &$21.9^{+2.1}_{-2.1}~~ 20.9^{+3.4}_{-2.1}$ &        $16.0^{+2.1}_{-2.1}$              \\ \smallskip

H${\beta}$(Red) (CMO) &  I($U$) (CMO) &8824-9033& Dec.2019--Jul.2020 &0.902~~~${\bf8.9^{+2.0}_{-5.0}}~~~7.9^{+3.0}_{-4.0}$ &$8.6^{+5.4}_{-2.7}~~ 10.6^{+3.8}_{-4.0}$&     $7.7^{+1.0}_{-3.0}$   \\ \smallskip

H${\beta}$(Red) (WIRO) &  I($U$) (CMO) &8749-9001& Sep.2019--May 2020 &0.918~~~${\bf6.1^{+1.6}_{-2.2}}~~~~ 5.7^{+2.4}_{-2.4}$ & $5.1^{+6.3}_{-2.4}~~ 9.4^{+1.6}_{-0.6}$&   $6.7^{+1.0}_{-1.2}$                   \\ \smallskip

H${\alpha}$ (CMO+WIRO ) &  I($U$) (CMO+WIRO) &8438-9033& Nov.2018--Jul.2020 &0.895~~${\bf21.2^{+2.8}_{-1.8}}~~ 18.0^{-3.4}_{-1.4}$ &$20.0^{+2.5}_{-3.9}~~ 20.8^{+3.0}_{-2.7}$& $19.7^{+2.8}_{-0.7}$                    \\ \smallskip

H${\beta}$ (CMO+WIRO) &  I($U$) (CMO+WIRO) &8438-9033& Nov.2018--Jul.2020 & 0.931~~${\bf17.5^{+1.9}_{-1.7}}~~ 15.9^{+2.0}_{-1.4}$ &$17.3^{+2.2}_{-2.3}~~ 15.5^{+4.0}_{-2.1}$&  $16.6^{+0.5}_{-0.2}$                   \\

\hline
\end{tabular}
 \label{Tab1}
\end{table*}

\section{Broad emission line profile variations: implications for the BLR dynamics and the black hole mass}
\begin{figure}
	\includegraphics[width=\columnwidth]{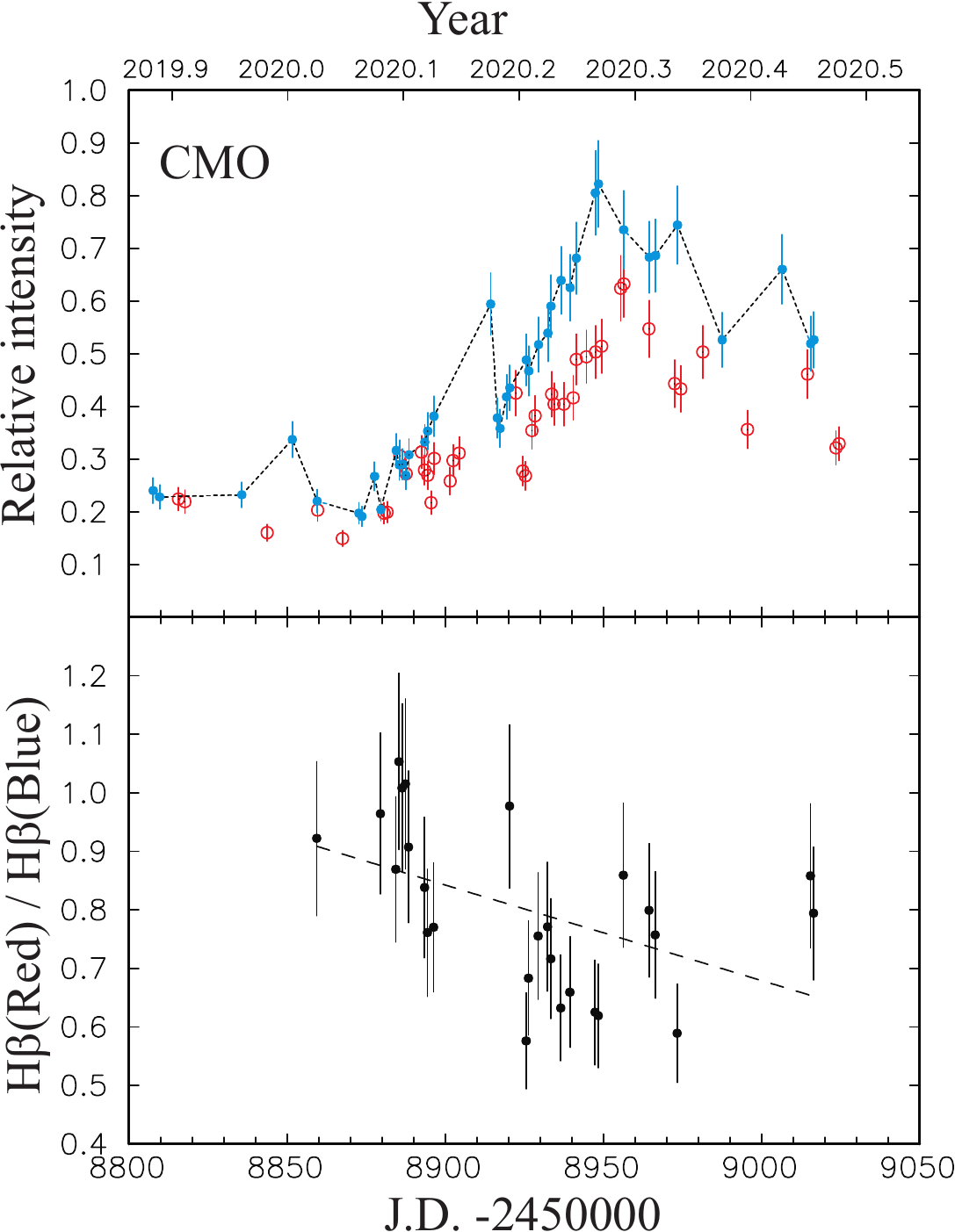}
    \caption{Top panel: Light curves for relative intensity   I(H{$\beta$}(Blue))   (dots) and I(H{$\beta$}(Red))  (open circles) shifted by -17 and -9 days, respectively (The scale was not reduced as at the Fig.~\ref{fig12} and values are the same as in Table~\ref{Tab2d}). Bottom panel: variations of the   I(H{$\beta$}(Red)) / ([H{$\beta$}(Blue)) ratio for the values shifted as it is shown on the top panel. We use just those interpolated values of  I(H{$\beta$}(Red)) which are farther not more than 2 days from the nearest observation.}    
     \label{hb-br}
\end{figure}

A large variation in H$\beta$ intensity was accompanied by a significant change in the line profile. The examples of H$\beta$ profiles for the maximum and minimum states are presented in Fig.~\ref{fig4} (CMO) and Fig.~\ref{fig5} (WIRO). The profiles of H$\beta$ in the high state are asymmetric and double-peaked with the blue peak dominating. The asymmetry was stronger in the maximum state during April--May 2020. This fact is clearly seen in Fig.~\ref{hb-br} where the light curves of H$\beta$(Blue) and H$\beta$(Red) are shown shifted by $-$17 and $-$9 days, respectively. There are also present the variations of the ratio of H$\beta$(Red) / H$\beta$(Blue) to show variability of the line asymmetry.

We have found not only considerable variations in the H$\beta$, H$\alpha$, H$\gamma$, \ion{He}{ii}$\lambda$4686 lines, but the appearance of strong  Balmer continuum in the high state, too (see Fig.~\ref{fig3}), which  is a typical feature for the CL AGN \citep[see e.g.,][]{Shappee2014, Oknyansky2020c}. We have also detected prominent coronal lines and possible variations of these lines that is also indicative of a CL event. The details on the variability of the coronal lines will be given in our future publication. In this work, we report on a strong outburst in NGC~3516 and on the recovering of the object back to the high state, which is similar to the previous CL event in 2016 \citep{Oknyansky2020c}.

The time delay of about $\tau\sim19$ days (taking into account the 2 days delay for the optical from X-ray variations) found for H$\beta$  ($R_{BLR}=c\tau$ is the emissivity-weighted radius of the BLR) and the full-width-half-maximum (FWHM) of the line rms-profile of about 4700 $km s^{-1}$ can be used to estimate the virial black hole (BH) mass:
\begin{equation}
\label{eqn1}
M_{BH} = f\frac{R_{BLR}\Delta V^2}{G},
\end{equation}
where $G$ is the gravitational constant and $f$ is the virial factor determined by the geometry and kinematics of the BLR \citep[e.g.,][]{Peterson2004}. Adopting the recent value for the $f$ factor from \cite{Woo2015} who obtained $\log f = 0.05\pm $0.12 for the FWHM-based $M_{BH}$ estimates, we obtain a mass of  $\sim 5\times10^7 M_\odot$ for the central BH.  This value is generally consistent with similar virial H$\beta$-based RM estimates published by \cite{Peterson2004} based on data from \cite{wa93} of 4.27$\pm1.46\times10^7 M_\odot$ ;
\cite{de10} of 3.17$^{+0.28}_{-0.42}\times10^7 M_\odot$ ;
\cite{dr18} of 4.27$^{+1.48}_{-1.11}\times10^7 M_\odot$;
\cite{sh19} of 4.73$\pm{1.40}\times10^7 M_\odot$ ;
\cite{Feng2020} of 2.4$^{+0.7}_{-0.3}\times10^7 M_\odot$ ;
and with the mass of $(1-5)\times10^7 M_\odot$ obtained using the variability of the Fe K$\alpha$ X-ray line reported by \cite{Iwasawa2004}. The high precision of the above values reflects measurement errors, while systematic errors due to issues of choice of measurement algorithm, calibration of the $f$ factor and its deviation from the average \citep[e.g.,][]{Pancoast2014b}, as well as the choice of method for measuring the velocity $\Delta V$
\citep[e.g.,][]{Bonta2020}, make the masses much more uncertain.

We want to point out that the very recent results of \cite{Feng2020}
were based on a six-month campaign from November 2018 to May 2019 that covered the epoch of the first and fainter of the two brightenings we also observed.  They found time lags generally smaller than ours ($\sim$7.5 days in the case of H$\beta$), which is not unexpected given the lower average luminosity; different RM campaigns listed above have found H$\beta$ time lags of $\sim$ 1-3 weeks.  Our time lags are dominated by the much larger variation in the second, larger brightening and are closer to the longer delays reported. Both our measurements and those of \cite{Feng2020} find the same dynamics, with similarly longer time delays for the blue side of the H$\beta$ profile, but the difference is that during the second brightening the blue side strengthened much more than the red side ( Fig.~\ref{hb-br}). The blueshifted emission region on the far side of the BLR would then perhaps be more clearly seen than red-shifted gas on the closer side of the BLR, due to opacity and asymmetric emission preferentially back toward the continuum source.


These considerations suggest radial infall rather than orbital motion. Orbital motion of gas in a flattened disk has often been preferred in other objects with double-peaked profiles that show symmetric velocity-resolved time lags.  The profile asymmetries and much longer time lags on the blueside suggest an alternative to a flattened disk with orbiting clouds. Let us consider a simple model with two-sided symmetrical radial inflows with effective  luminosity-weighed centres located at the distance $R_{in}$ from the central BH in the direction that forms an angle $\delta$ to the line of sight.
In this case, the time delay (luminosity-weighed) for the emission region responsible for the blue peak will be $\tau_{bl}c =R_{in}(1+\cos{\delta})$ and for the red, $\tau_{red}c =R_{in}(1-\cos{\delta})$.
 The measured values of $\tau_{bl}$=17 days and $\tau_{red}$= 9 days must be increased by 2 days corresponding to the optical from X-ray delay. Then  we can estimate $\delta$ as $\sim75^{\circ}$ and $R\sim15$ l.d. The velocity in the maximum in the blue peak of H$\beta$ is $\sim$1700 km s$^{-1}$ corresponding to the radial inflow velocity $V_{r}\sim$6400 km s$^{-1}$.

If we assume that this velocity relates to free fall in the gravitation field of the central BH, then by analogy from Equation~\ref{eqn1} we get

\begin{equation}
\label{eqn2}
 M_{BH} = \frac{R_{in} V_{r}^2}{2G}.
\end{equation}

According to Equation~\ref{eqn2}, the BH has a mass of $\sim 5\times10^7 M_\odot$, that agrees well with the estimates above.  This is promising in that perhaps the virial equation, commonly adopted for mass estimation when only single-epoch spectra are available, does not provide wildly inconsistent values compared with a likely radial flow producing an extremely asymmetric profile.

\section{Evolution of the X-ray spectrum}

To trace the evolution of the X-ray spectrum of NGC~3516 as a function of luminosity, we performed analysis of three {\it Swift}/XRT spectra  with large enough exposure (to have about similar signal/noise ratio in each spectrum)  obtained in very different states. Particularly, the source was observed in the low  state with a flux of $F_{\rm 0.5-10 keV}=7.3\times10^{-12}$ erg s$^{-1}$ cm$^{-2}$ on July 11, 2014 (ObsID 00080749004), whereas the maximal flux $F_{\rm 0.5-10 keV}=9.8\times10^{-11}$ erg s$^{-1}$ cm$^{-2}$ was observed around April 1, 2020 (ObsID 00035462024).

For comparison we also analysed one set of observational data obtained one month after the  maximum when the source was in an intermediate   flux state with $F_{\rm 0.5-10 keV}=4.7\times10^{-11}$ erg s$^{-1}$ cm$^{-2}$ on May 2, 2020 (ObsID 00035462032).  
Spectral analysis has been done using {\sc xspec} package \citep{Arnaud1996} and applying W-statistics \citep{Wachter1979} after the spectra were binned to have at least one count in each energy channel.

\begin{figure*}
\centering
\includegraphics[scale=0.9,angle=0,trim=0 0 0 0]{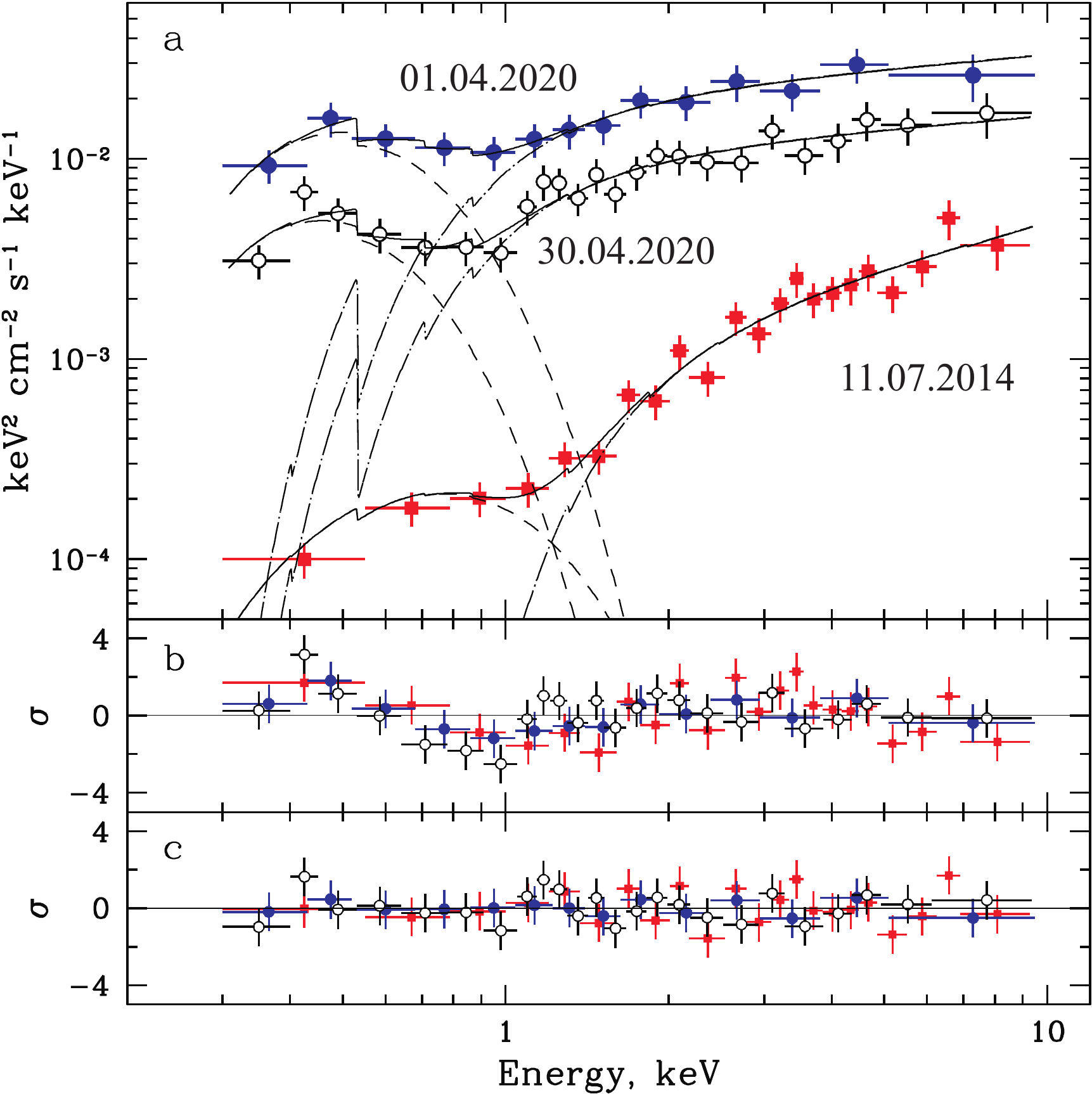}

\caption{(a) Unfolded X-ray spectra of NGC~3516 obtained with {\it Swift}/XRT in different states. Spectra in high (ObsID 00035462024), intermediate (ObsID 00035462032) and low (ObsID 00080749004) flux states are shown with solid circles, open circles and solid squares, respectively. Solid line represent the best-fitting model consisting of the absorbed power-law and black body (also shown with dash-dotted and dashed lines, respectively).
}\label{fig:spe}
\end{figure*}

We started our spectral analysis with the simplest model consisting of a power law modified by photoelectric absorption ({\sc phabs$\times$po} models in {\sc xspec}). The residuals for this model shown in Fig.~\ref{fig:spe}b clearly demonstrate an unacceptable fit. Particularly, either a broad absorption between 0.6 and 1.1 keV, or additional soft emission component below 1 keV, is required.  Applying this simplified model to all available spectra we derived the equivalent hydrogen column density consistent with zero and photon index ranging from 0.6 to 1.6 for the lowest and highest states, respectively.

As the next step we modified our initial model adding a soft black body component. To take into account the Galactic photoelectric absorption we introduced the second {\sc phabs} component and fixed its value at $0.03\times10^{22}$~cm$^{-2}$ \citep{Collaboration2016}. Therefore, the final best-fit model for all the three {\it Swift}/XRT spectra is {\sc phabs$\times$(bb+phabs$\times$po)}. The second {\sc phabs} component represents  intrinsic absorption and was left free to vary. The resulting spectra and corresponding spectral parameters are shown in Fig.~\ref{fig:spe} and Table~\ref{tab:spe}, respectively.


As it is seen from Fig.~\ref{fig:spe} the change between high and post-maximum intermediate states is about the same for all the energy regions. Such type of variability can be due to either intrinsic continuum variations or `colorless` variability originating from the passage of Compton-thick blobs of gas across the line-of-sight. Taking into account the dropout of the X-ray and UV correlation for the same dates makes the second explanation more plausible.

\begin{figure*}

\includegraphics[scale=1.3,angle=0,trim=0 0 0 0]{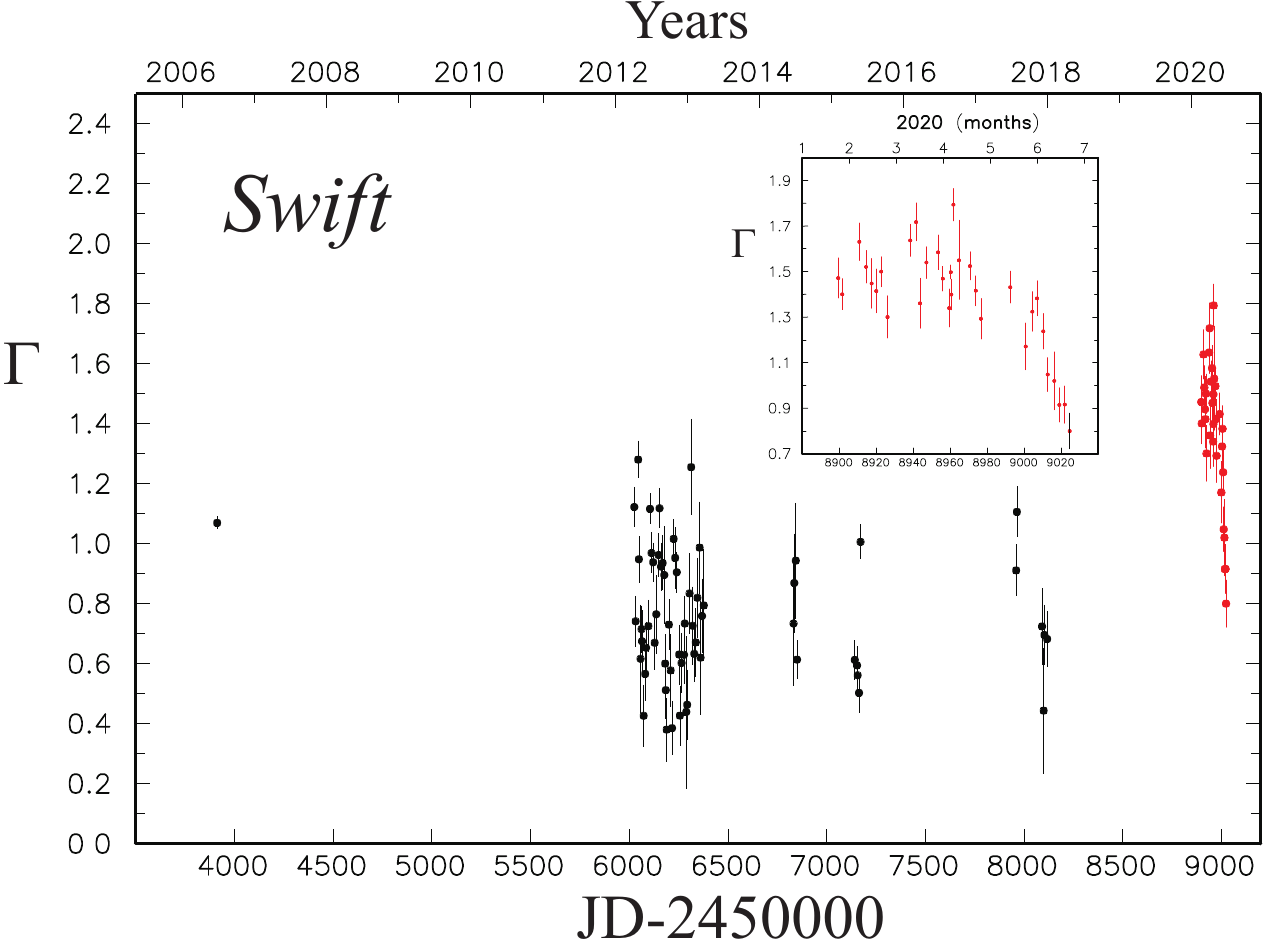}
    \caption{Evolution of the photon index value, calculated by assuming a simple absorbed power-law model, based on the {\it Swift}/XRT data. The inset shows results obtained during first half of 2020. (See details in the text).} 
       \label{fi}
\end{figure*}

A significant change of about 2 orders of magnitude of the soft X-ray from the minimal state to the high and intermediate ones rejects the possibility to explain this soft component as thermal radiation from the host galaxy and so the preference should be given to the models which predict strong variability of the soft excess \citep[e.g.,][]{Gardner2017}.

\begin{table}
  \noindent
\caption{Best-fit parameters for a {\sc phabs$\times$(bb+phabs$\times$po)} model obtained for three {\it Swift}/XRT observations. All errors are reported at 1$\sigma$ confidence level.}
\label{tab:spe}
\begin{tabular}{ccc}
\hline
\hline
Parameter &     Units                   &           Value \\
\hline
\multicolumn{3}{c}{High state (ObsID 00035462024)}  \\
\hline
$N_{\rm H}$1 & $10^{22}\text{ cm}^{-2}$ &  $0.03^a$ \\
$kT^{b}$         &          keV         &  $0.11\pm0.02$ \\
Norm$_{\rm bb}$  &  $L_{39}/D^2_{\rm 10 kpc}$  &  $(4.6\pm0.8)\times10^{-4}$ \\
$N_{\rm H}$2 & $10^{22}\text{ cm}^{-2}$ &  $0.3\pm0.2$ \\
Power-law ph. index     &               &   $1.8\pm0.3$   \\
Power-law norm.      &                  &   $(1.9\pm0.8)\times10^{-2}$   \\
Flux$^c$ & $10^{-11}$ erg s$^{-1}$ cm$^{-2}$ & $9.8\pm0.2$\\
\hline
\multicolumn{3}{c}{Intermediate state (ObsID 00035462032)}  \\
\hline
$N_{\rm H}$1 & $10^{22}\text{ cm}^{-2}$ &  $0.03^a$ \\
$kT^{b}$         &          keV         &  $0.10\pm0.01$ \\
Norm$_{\rm bb}$  &  $L_{39}/D^2_{\rm 10 kpc}$  &  $(1.8\pm0.3)\times10^{-4}$ \\
$N_{\rm H}$2 & $10^{22}\text{ cm}^{-2}$ &  $0.32\pm0.15$ \\
Power-law ph. index     &               &   $1.8\pm0.2$   \\
Power-law norm.      &                  &   $(1.0\pm0.2)\times10^{-2}$   \\
Flux$^c$ & $10^{-11}$ erg s$^{-1}$ cm$^{-2}$ & $4.7\pm0.2$\\
\hline
\multicolumn{3}{c}{Low state (ObsID 00080749004)}  \\
\hline
$N_{\rm H}$1 & $10^{22}\text{ cm}^{-2}$ &  $0.03^a$ \\
$kT^{b}$         &          keV         &  $0.18\pm0.03$ \\
Norm$_{\rm bb}$  &  $L_{39}/D^2_{\rm 10 kpc}$  &  $(6.4\pm0.8)\times10^{-6}$ \\
$N_{\rm H}$2 & $10^{22}\text{ cm}^{-2}$ &  $1.3\pm0.5$ \\
Power-law ph. index     &               &   $1.2\pm0.3$   \\
Power-law norm.      &                  &   $(8.5\pm3.0)\times10^{-4}$   \\
Flux$^c$ & $10^{-12}$ erg s$^{-1}$ cm$^{-2}$ & $7.3\pm0.3$\\
\hline
\end{tabular}
	\begin{tablenotes}
	    \item $^{a}$ Fixed.
        \item $^{b}$ Black body temperature.
        \item $^{c}$ Observed flux in the 0.5--10 keV energy band.
    \end{tablenotes}
\end{table}

In addition to detailed fits of these three epochs, we measured the photon index power law $\Gamma$ more generally for the entire Swift XRT data set for NGC~3516  for the simple absorbed power law model. Fig.~\ref{fi} shows generally small values of $\Gamma$ when the X-ray flux is low, but there is a significant increase during the 2020 brightening, although falling off at the end.  This is consistent with an intrinsic version
of the correlation between $\Gamma$ and Eddington ratio previously reported \citep[e.g.,][]{Risaliti2009}.  These relatively small values of $\Gamma$ are suggestive of low Eddington fractions associated
with the transition between different Seyfert classes as seen in NGC~3516.

\section{Discussion and conclusions}

NGC~3516 is one of the typical examples of an AGN  for which spectra range from type Sy1.2 to type Sy1.9 and back for the same object over long time scales (years) or sometimes shorter time scales (months). Using our spectroscopy and multi-wavelength photometry, we have shown that NGC~3516, after several years being in a low state, is recently again in a high state with strong broad emission lines  and the highest flux level of its continuum. These dramatic changes can be identified as a new CL event for the object.

We have found  delays between continuum variations at different wavelengths (X-ray, UV, optical), as well as delays between continuum and broad emission lines variability using 3 different statistical methods.  The wavelength-dependent time-delays between the continuum at different wavelengths are predicted in the  most common standard model with the thermal reprocessing hypothesis in AGN, where extreme ultraviolet/X-ray photons are reprocessed by the accretion disc (AD) into optical/UV photons \citep[see e.g.,]{Cackett2007}.  In this  model it is  assumed that the X-ray emitter is located on the rotating axis of the   supermassive black hole (SMBH) like a ``lamppost''  and its height $H{_x}$  from the SMBH is much smaller than $R$ -- distance to the  location were the UV/optical radiation arises in AD. Our inference that the UV variations lag behind the X-ray ones by $\sim$ 1.6 days  (which is in agreement with previously published results \cite{bu17, nd16}) is in contradiction with the lamppost-type X-ray reprocessing model, that is a common problem for other typical Seyfert galaxies (which are mostly also known as CL AGNs) including NGC~4151, NGC~5548, NGC~4051, NGC~2617, NGC~1566,  and Mrk 335  \citep[see e.g.,][]{Cackett2007,  bu17,  Edelson2017, McHardy2014, Shappee2014, Oknyansky2017a, Oknyansky2020c}. For some gravitationally lensed quasars the sizes of ADs measured from the microlensing data are also significantly larger than those predicted by the most commonly adopted X-ray reprocessing model \citep[e.g.,][]{Morgan2010}. First of all, the observed luminosity (estimated Eddington ratios) has to be much higher for the size of the accretion disk determined by the observed time lag and microlensing estimations  \citep{Cackett2007}. If to take $H{_x}\sim R$ which is required to explain the observed delays, then the necessity to explain sufficient primary X-ray radiation from such an X-ray corona makes this possibility highly unrealistic \citep{nd16}. Two possible alternative models were discussed by \cite{nd16} (see details therein) to explain such big lags in UV/optical variation:  truncated AD in analogy with black-hole binaries (e.g, \citet{Done2007})  and  the model with a soft excess region at the inner edge of an accretion disk which completely hides the hard X-ray corona \citep{Gardner2017}.  The second model was also used to explain the time delays in NGC~4151 \citep{Edelson2017}.  However, \cite{nd16} prefer the truncated AD model for NGC~3516 on the grounds that soft X-ray variations are not detected. Our results indicate strong variability of the soft component, which can not be associated with the host galactic thermal emission. So, our results do not reject the second model.
The big values for UV/optical delay can also be explained assuming that some part of the flux arises not from the AD but relates to the diffuse continuum radiated by the same gas that emits the broad emission lines. \citep[see e.g.,][]{Goad2019}. For example, a delay in $U$ of about 0.4 days relative to $B$ found by us can be explained by the Balmer continuum input to the band, which is clearly seen in our spectra (see Fig.~\ref{fig3}).

 We have found a delay of about 17 days for the variation in H$\beta$ and predicted the variations of the line on the base of optical photometry. We have found the time delays slightly decreased for the Balmer lines from maximal for H$\alpha$ to minimal for  H$\gamma$.  The same result was found for some other AGNs \citep[e.g.,][]{Feng2021} and can be due to the decrease of the optical depth along the H$\alpha$, H$\beta$ and H$\gamma$ lines. The time delay for He~II was found to be smaller than for all the Balmer lines, as is the case in many other AGNs \citep[e.g.,][]{Grier2013,Peterson2014a} which can be due to the fact that the emission region of higher ionisation is located closer to the central source of the UV ionisation flux. These time delays allow us to estimate the spacial scales corresponding to the emission regions in the BLR.

Double-peaked broad emission line profiles are seen in many AGNs and were intensively investigated both in observational and theoretical ways \citep[see e.g.,][]{strateva2003, Du2018}.  In NGC~3516 the double emission line profile was seen in previous spectra \citep{sh19} but in our spectra taken in 2019-2020 the emission peaks in the blue and red wings of H$\beta$ are more prominent and the blue one is stronger than the red. \cite{strateva2003} argued that the situation when the blue peak is stronger than the red could be explained by a circular disc emission model. 
At the same time, the largest delay for the negative velocities in the H$\beta$ emission may indicate inflowing kinematics of the gas clouds, inconsistent with disc orbital motions. The observed characteristics are about the same as were found in 2007 \citep{Denney2009}, when the object was in the high state and probably are different from those in 2012 \citep{dr18} when the object was in the low state. This does not necessarily mean that the kinematics of the clouds have changed in such a short time. This may be due to the fact that when the luminosity of the object changes, the size of the region which effectively radiates in broad emission lines varies, too, and the kinematics of the clouds can depend on the distance to the central source.  The profile of H$\beta$ in Mar.-Jul. 2020 is similar to that observed in 2007.  That is in agreement with our conclusion that the object has changed its type back to Sy1.5 again.  The CL event observed in 2020 is similar to a previous one that occurred about 4.5 years before \citep{Oknyansky2019a}. This time interval is in agreement with regularity in optical variations of the object found  by \cite{Kovacevic2020}. 

Dramatic changes of the optical emission lines followed a strong outburst in the X-rays and UV.
On 1 Apr. 2020 the X-ray flux reached a maximal level (since Swift observations began in 2006) of about 9.8 $\times$ 10$^{-11}$ ergs cm $^{-2}$ s$^{-1}$ that is about 22 times bigger than in the observed minimum on 27 Jun. 2014. This amplitude is compatible with what is seen (20--50 times) in other highly variable CL AGNs \citep[see e.g.,][]{Oknyansky2019b,Zetzl2018}.  
The amplitude of the UV outburst was about 3 magnitudes, but the maximum was about 3 weeks later than in the X-ray. Variations of X-ray and UV/optical radiation were different during the second half of April  -- June 2020.  At the first date of $Swift$ observations in 2006 (see Fig.~\ref{fig1}) a high level of UV brightness was observed, but rather intermediate one of X-ray flux, though they were mostly well correlated during 2012--2020. These events of  colorless variations of X-ray flux which are not correlated with UV/optical variability might be suggested to be a consequence of Compton-thick clumps of gas crossing the line-of-sight and temporarily obscuring the X-ray source which is significantly smaller than the UV/optical one \citep[see e.g., for NGC~2617][]{Oknyansky2017a}.
A similar event was observed during the previous high state of NGC~3516 in 2009 and the same explanation was invoked by \cite{Turner2011}.

The observed change in asymmetry of H$\beta$ in April - May 2020 can also be associated with some absorption (coinciding in time with the proposed absorption in X-ray), which weakened ionizing radiation in the direction of the emission region of the red peak in the line. The presence of absorption predominantly on one side of the AD indicates a certain asymmetry in the distribution of absorbing dense clouds, which can be used for selecting a possible model for CL events.

We offer as an alternative a straightforward explanation of the changes in H$\beta$ profile
asymmetries and velocity-resolved time lags observing in NGC~3516.
A BLR with a flattened geometry seen more edge-on than face-on will have
the near-side respond with shorter time lags than the far side.
Moreover, we may expect the side of the line-emitting gas facing the continuum
source to emit more strongly and preferentially back in the direction
of the continuum, and thus be seen more strongly from our perspective
\citep{Pancoast2014b}.
If we consider the NGC~3516 BLR to be dominated by radial motions over
orbital motions, then we have an understandable picture.
In 2007, we see excess emission on the blue side of the H$\beta$ profile
and the longest time lags indicating inflow \citep{de10}.
In 2012, we see the peak emission shifted redward to near zero velocities,
and larger time lags near zero and slightly redshifted velocities,
indicating some dynamics more similar to outflow \citep{dr18}.
In 2019-2020, after an extended low state, we again see H$\beta$ time lags
longest on the blue side as well excess emission on blue side,
indicating inflow again (e.g. see also, \citet{Feng2020}).
In an object like NGC~3516 with a likely edge-on BLR dominated by radial
flow as demonstrated by RM, the H$\beta$ profile asymmetries track
the velocity-resolved time lag distributuions.  In some sense, we
can perhaps regard this object's BLR as now understood, and we
can model it in the future.

There are several physical processes which are considered as a possible reason for such dramatic changes of the luminosity in outburst and changing the Seyfert type: variable absorption along the line of sight, various types of instabilities in the accretion disk, and tidal destruction of stars (TDE).
The strong UV flux following the change can sublimate the dust in a biconical hollow outflow \citep[see][]{oknyansky2015} and, as a result, we may see the BLR without obscuration. So, in some not-too-long time after the outburst the dust can reform and the object will be seen as a type 2 again, the broad lines hidden from our line of sight. The most significant problem concerning TDEs is too low of a cadence of such accidents, which cannot explain the amount of observed CL events. An alternative idea involving the tidal stripping of stars \citep{Campana15, iv06} could lead to more frequent (than TDEs) and recurrent events of dramatic changes in the AGN accretion rate. Recurrent CL events in AGN can find a natural explanation, as in models with instability in the AD as well as in the model with tidal stripping of stars.   Recently \cite{Wang2020}  also suggested a novel explanation for CL events, and for their recurrences in individual objects, by invoking the case of close binary black holes with a low mass ratio. For the last two models, an asymmetrical distribution of gas clouds near the AD may be expected.


Our observations of NGC~3516 during this latest CL event will not solve the problem of CL events more generally, but they do add to the phenomenology that must be addressed by proposed explanations.  More references on the topic can be found in discussions by \cite{MacLeod2019, Oknyansky2017a, Oknyansky2019, Runnoe2016, Ruan2019}.

\section*{Acknowledgements}
P. Du acknowledges financial support from the National Science Foundation of China (12022301, 11873048 and 11991051) and from the Strategic Priority Research Program of the CAS (XDB23010400). J.-M. Wang acknowledges financial support from the National Science Foundation of China (11991054 and 11833008), from the National Key R$\&$D Program of China (2016YFA0400701), from the Key Research Program of Frontier Sciences of the Chinese Academy of Sciences (CAS; QYZDJ-SSW-SLH007), and from the CAS Key Research Program (KJZD-EW-M06). 

Part of this work was supported by the M.V. Lomonosov Moscow State University Program of Development (RC600 \& TDS). {EOM, SAP, NIS, AVD acknowledge the support by the Interdisciplinary
Scientific and Educational School of Moscow University 'Fundamental and Applied Space Research'. EOM, NIS, AVD, AMT, SGJ are also supported by the RSF grant 17-12-01241. We thank the $Swift$ team for carrying out our ToO  requests. We are grateful to R. Oknyansky for help with  ICCF and JAVELIN code installation and adoption.  We thank WIRO engineers James Weger and J. Conrad Vogel for their observatory support.  This work is supported by the National Science Foundation under REU grant AST1852289 and PAARE grant AST 1559559.
Michael Brotherton enjoyed support from the Chinese Academy of Sciences President’s International Fellowship Initiative, Grant No. 2018VMA0005.
We also acknowledge support from a University of Wyoming Science Initiative Faculty Innovation Seed Grant.
T.E. Zastrocky acknowledges support from NSF Grant 1003588R.

\section*{Data availability}
The data underlying this article are available in the article and in its online supplementary material.



\bibliographystyle{mnras}
\expandafter\ifx\csname natexlab\endcsname\relax\def\natexlab#1{#1}\fi
\bibliography{ngc3516} 

\bsp	
\label{lastpage}
\end{document}